\documentclass[12pt]{imsart}
\RequirePackage[OT1]{fontenc}
\RequirePackage{amsthm,amsmath}
\RequirePackage[numbers]{natbib}
\RequirePackage[colorlinks,citecolor=blue,urlcolor=blue]{hyperref}

\usepackage{graphicx,amsmath,amsthm,amsfonts,amssymb}
\usepackage{booktabs, rotating, float}
\usepackage[hmargin=3cm,vmargin=3cm]{geometry}
\usepackage{mathtools}
\mathtoolsset{showonlyrefs}
\usepackage{enumerate}
\usepackage{color}
\usepackage{listings}
\usepackage{epstopdf}
\usepackage{makecell}
\usepackage{tikz}
\usetikzlibrary{decorations.pathreplacing}
\usepackage{array}
\usepackage{caption}
\usepackage{subcaption}
\usepackage{soul}
\usepackage{hyperref}
\usetikzlibrary{calc}
\theoremstyle{plain}
\newtheorem{thm}{Theorem}[section]
\newtheorem{cor}{Corollary}[section]

\newtheorem{prop}{Proposition}[section]
\newtheorem{lemma}{Lemma}[section]
\newtheorem{assumption}{Assumption}[section]
\theoremstyle{definition}
\newtheorem{defn}{Definition}[section]

\newtheorem{eg}{Example}[section]
\newtheorem{fact}{Fact}[section]
\newtheorem{observe}{Observation}[section]

\numberwithin{equation}{section}

\newcommand{\rpm}{\sbox0{$1$}\sbox2{$\scriptstyle\pm$}
  \raise\dimexpr(\ht0-\ht2)/2\relax\box2 }

\newcommand{\R}{\mathbb{R}}

\tikzstyle{nd} = [anchor=base, inner sep=0pt]
\tikzstyle{ndpic} = [remember picture, baseline, every node/.style={nd}]

\def\beq{\begin{equation}}
\def\eeq{\end{equation}}
\def\ba{\begin{enumerate}[(a)]}
\def\bei{\begin{enumerate}[(i)]}
\def\be{\begin{enumerate}[(1)]}
\def\ee{\end{enumerate}}
\def\bi{\begin{itemize}}
\def\ei{\end{itemize}}
\def\beg{\begin{eg}}
\def\eeg{\end{eg}}
\def\bd{\begin{defn}}
\def\ed{\end{defn}}
\def\bt{\begin{thm}}
\def\et{\end{thm}}
\def\bl{\begin{lemma}}
\def\el{\end{lemma}}
\def\bfac{\begin{fact}}
\def\efac{\end{fact}}

\def\bc{\begin{cor}}
\def\ec{\end{cor}}
\def\bp{\begin{prop}}
\def\ep{\end{prop}}
\def\bo{\begin{observe}}
\def\eo{\end{observe}}
\def\bas{\begin{assumption}}
\def\eas{\end{assumption}}
\def\RR{\mathbb{R}}

\def\EE{\mathbb{E}}

\def\NN{\mathbb{N}}

\def\PP{\mathbb{P}}
\def\SS{\mathbb{S}}
\def\Sn{\mathbb{S}_n}
\def\Rn{\mathbb{R}_n}
\def\tRn{\tilde{\mathbb{R}}_n}
\def\Yl{Y_n^{loc}}
\def\Hl{H_n^{loc}}
\def\tHl{\tilde{H}_n^{loc}}
\def\ii{\item}

\def\beg{\begin{eg}}
\def\eeg{\end{eg}}
\def\t{\mathrm{T}^{(1)}_{u,v}}
\def\tt{\mathrm{T}^{(2)}_{u,v}}

\def\hm{\hat{\mu}_n}

\def\p{\thinspace}

\makeindex

\numberwithin{equation}{section}
\numberwithin{table}{section}
\newtheorem{remark}{Remark}

\startlocaldefs
\endlocaldefs

\begin{document}

\begin{frontmatter}

\title{On Univariate Convex Regression}
\runtitle{On Univariate Convex Regression}


\begin{aug}
  \author{\fnms{Promit}  \snm{Ghosal}\ead[label=e1]{pg2475@columbia.edu}} \&
  \author{\fnms{Bodhisattva} \snm{Sen}\corref{}\thanksref{t2}\ead[label=e2]{bodhi@stat.columbia.edu}\vspace{0.1in} \\ {\it Columbia University}}
  
   \affiliation{Columbia University}

  \thankstext{t2}{Supported by NSF grant DMS-1150435}

  \runauthor{Ghosal \& Sen}

  \address{1255 Amsterdam Avenue, New York, NY 10027\\
          \printead{e1,e2}}

\end{aug}

\begin{abstract}
\noindent We find the local rate of convergence of the least squares estimator (LSE) of a one dimensional convex regression function when (a) a certain number of derivatives vanish at the point of interest, and (b) the true regression function is locally affine. In each case we derive the limiting distribution of the LSE and its derivative. The pointwise limiting distributions depend on the second and third derivatives at 0 of the ``invelope function'' of the integral of a two-sided Brownian motion with polynomial drifts. We also investigate the inconsistency of the LSE and the unboundedness of its derivative at the boundary of the domain of the covariate space.  An estimator of the argmin of the convex regression function is proposed and its asymptotic distribution is derived. Further, we present some new results on the characterization of the convex LSE that may be of independent interest.
\end{abstract}




\end{frontmatter}









\section{Introduction}
Consider the regression model 
\[Z=\mu(X)+\varepsilon,\]
where $X$ is uniformly distributed on $[0,1]$, $\varepsilon$ is the (unobserved) mean zero error independent of $X$ and $\mu:[0,1]\to \RR$ is an unknown \emph{convex} function.  Given i.i.d.~observations $(X_1,Z_1),(X_2,Z_2),\ldots ,(X_n,Z_n)$ from such a model the goal is to estimate the unknown regression function $\mu$. We consider the least squares estimator (LSE) $\hm$ of $\mu$ defined as any convex function that minimizes the $\mathcal{L}_2$ norm
\begin{equation}\label{eq:LS}
\left(\sum_{i=1}^n\left(Z_i-\psi(X_i)\right)^2\right)^{1/2}
\end{equation}
among all convex functions $\psi$ defined on the interval $[0,1]$. Note that the computation of the LSE reduces to solving a quadratic program with $(n-2)$ linear constraints; see e.g.,~\citep{FM89}. 

Estimation of a convex/concave regression function has a long history in statistics. Least square estimation of a concave regression function was first proposed by Hildreth~\citep{HLL54} for estimation of production functions and Engel curves. The consistency of the least squares concave regression estimator was first established in~\citep{HANPLG76}. The pointwise rate of convergence of the LSE  in convex regression, at an interior point $x_0 \in (0,1)$, was studied in~\citep{MAMMEN1991} under a uniform fixed design setting. Pointwise limiting distribution of the LSE, under the assumption that $\mu''(x_0) \ne 0$, was derived by \citep{GJW2001b}, again under a fixed design setting. 


In this paper we study the rate of convergence and the asymptotic distribution of $\hm(x_0)$ under the following two scenarios: 
\begin{itemize}
	\item[(a)] the $k$-th derivative $\mu^{(k)}(x_0) = 0$ for $k = 2, 3, \ldots, r - 1$, and $\mu^{(r)}(x_0) \ne 0$, where $r$ is an integer greater than 2; 
	
	\item[(b)] there exists an interval around $x_0$ such that $\mu$ is affine in that interval.
\end{itemize}
	We show that under scenario (a) the estimator $\hm(x_0)$, properly normalized, converges to a non-degenerate limit at the rate $n^{-r/(2 r +1)}$. We also show that $\hm'(x_0)$ converges at the rate $n^{-(r-1)/(2 r +1)}$. Under scenario (b), we show that both $\hm(x_0)$ and $\hm'(x_0)$ converge to non-degenerate limits at the rate $n^{-1/2}$.

Moreover, in this paper, we study the behavior of the LSE $\hm$ at the boundary of the domain of the predictor (i.e., at 0 and 1), and establish the inconsistency of $\hm$ at the boundary; we also show that the derivative of $\hm$ at the boundary is unbounded.  In addition, we study the estimation of the argmin (argument of the minimum) of $\mu$ and find the asymptotic distribution of our proposed estimator (see \citep[Theorem~3.6]{MR2509075} for a related result in the case of log-concave density estimation). Further, we present some new results on the characterization of the convex LSE that may be of independent interest.

Although there has been some work investigating the local behavior of the convex LSE in the related problem of a convex density estimation (see e.g.,~\citep{GJW2001b},~\citep{FB2007}), not much is known in the case of a random design regression setting. This has been our main motivation in writing this paper, and indeed, our paper will try to fill this gap in the literature. In \citep[Theorem~2.1]{MR2509075}, one can find a result related to the scenario (a) above in the context of log-concave density estimation. During the process of writing this paper we discovered a recent related paper~\citep{CW15} that addresses the  convex regression problem for the fixed uniform design regression in scenario (b) above (and also for the density estimation problem). In~\citep{CW15} the authors study a stylized least squares-type regression estimator that is close but different from our convex LSE. Further,~\citep{CW15} does not study the behavior of the convex LSE under scenario (a). Moreover, the proof techniques employed in~\citep{CW15} to study scenario (b) are very different from ours:
~\citep{CW15} crucially uses an extended version of Marshall's lemma (see e.g,~\citep{DRW07},~\citep{BR08}) to study the rate of convergence of the convex LSE whereas we directly compare the convex LSE with the simple linear regression line fitted with the data points in the interval where $\mu$ is affine and establish that the supremum distance between these two fitted functions over the interval stochastically decays down to zero at a rate faster than $n^{-1/2}$. 

We organize the paper as follows. In Section~\ref{CharLSE}, we study the characterization of the convex LSE and present two useful results. In Section~\ref{LocalRate} we derive the local rate of convergence, at an interior point $x_0 \in (0,1)$, of the LSE $\hm$ in the two special cases: (a) when all the first $r-1$ ($r \ge 2$) derivatives of $\mu$ vanish at $x_0$, and (b) when $\mu$ is affine in a small neighborhood around $x_0$. Section~\ref{AsympDist} is devoted to the study of the pointwise asymptotic distributions, under scenarios (a) and (b). In Section~\ref{Boundary} we establish the inconsistency of $\hm$ at the boundary of the support (i.e., at 0 and 1); we also show that  the derivative of $\hm$ at the boundary is unbounded. Estimation of the argmin of $\mu$ is addressed in Section~\ref{Minima}. Appendix~\ref{Appendix} contains the proofs of some of the results stated in the paper.  
  
\section{Characterization and representation of the LSE}\label{CharLSE}
For notational simplicity let us denote by $(x_1,x_2,\ldots ,x_n)$ the ordered version of the sample $(X_1,X_2,\ldots ,X_n)$ and by $\mathbf{Y}:= (Y_1,Y_2,\ldots ,Y_n)$ the concomitant response vector. Let $\mathcal{K}$ be the set of all convex functions defined on $[0,1]$. Thus, 
\begin{equation}\label{eq:OptProb}
\hm \in \underset{\psi \in \mathcal{K}}{\arg \min} \;\phi_n(\psi)\quad \text{where} \quad \phi_n(\psi)=\sum_{i=1}^n\left(Y_i-\psi(x_i)\right)^2.
\end{equation}
Note that $\hm$ is only unique at the data points $x_i$'s. As most authors, we define our LSE to be the linear interpolant of $\{(x_i,\hm(x_i))\}_{i=1}^n$ on $[0,1]$. Let
\[\mathcal{K}_n:=\left\{\Psi = \left(\psi_1,\ldots ,\psi_n\right) \in \RR^n:  \frac{\psi_2-\psi_1}{x_2-x_1}\leq \ldots \leq \frac{\psi_n-\psi_{n-1}}{x_{n}-x_{n-1}}\right\}.\]
The optimization problem~\eqref{eq:OptProb} reduces to the following quadratic program with linear constraints:
 \[\hat{\Psi}_n=\underset{\Psi \in \mathcal{K}_n} {\arg \min }\; \sum_{i=1}^n\left(Y_i-\psi_i\right)^2.\]
Here ${\psi}_i$ is identified with $\psi(x_i)$, for $i=1,\ldots, n$. As $\mathcal{K}_n$ is a closed convex polyhedral cone $\hat{\Psi}_n$ is the projection of $\mathbf{Y}$ onto $\mathcal{K}_n$. 
\bl\label{CharLemma1}
Any LSE $\hm$, defined via~\eqref{eq:OptProb}, satisfies the following conditions:
\bei
 \ii $(\hm(x_1),\ldots ,\hm(x_n)) \in \mathcal{K}_n$,
 \ii $\sum_{i=1}^n\hm(x_i)(Y_i-\hm(x_i))= 0$,
 \ii $\sum_{i=1}^n(\psi(x_i)-\hm(x_i))(Y_i-\hm(x_i))\leq 0$, for any convex function $\psi:[0,1] \to \R$.
 \ee 
 \el 
 \begin{proof}
The proof is a direct consequence of the characterization of projection onto a closed convex cone; see e.g., \citep[Theorem 1.3.2]{RWD88}. 
 \end{proof}     
The above characterization has the following simplified form which is proved in \citep[Lemma 2.6]{GJW2001b}.
\bl\label{CharLemma2}
Let $\widehat{R}_{n,k}=\sum_{i=1}^k \hm(x_i)$ and $S_{n,k}=\sum_{i=1}^k Y_{i}$. Then $\hm \in \underset{\psi \in \mathcal{K}}{\arg \min}\; \phi_n(\psi)$ if and only if $\widehat{R}_{n,n}=S_{n,n}$ and
 \begin{align}\label{eq:charlemma2}
 \sum_{k=1}^{j-1}(\widehat{R}_{n,k} - S_{n,k}) \left(x_{k+1}-x_{k}\right)\left\{\begin{array}{ll}
 \geq 0, \quad j=2,3,\ldots ,n,\\
 =0, \quad \text{if $\hm$ has a kink at $x_{j}$ or $j=n$,}
 \end{array}\right.
 \end{align} 
where a kink (at $x_j$) signifies a change of slope of the linear interpolant of the $(x_i,\hm(x_i))$, for $i=1,\ldots, n$.
 \el
Next, we give a few simple consequences of the above two results.  
\bl\label{Char3}
\bei
\ii
The convex LSE $\hm$ satisfies $$\sum_{i=1}^n\left(Y_i-\hm(x_i)\right)=0, \qquad \sum_{i=1}^n x_i \left(Y_i-\hm(x_i)\right)=0.$$
\ii Suppose that $u<v$ be the two end-points of a affine part of $\hm$. Let us denote $u$ by $x_k$ for some $1\le k\le n$. Then,
\begin{align}
\sum_{i:u\leq x_i\leq v}\left(Y_i-\hm(x_i)\right)&\leq 0, \label{eq:imp1}\\
\sum_{i:u\leq x_i\leq v}x_i\left(Y_i-\hm(x_i)\right)&\leq 0, \label{eq:imp2}\\
Y_k-\hm(x_k)&\leq 0, \label{eq:imp5}\\
\sum_{i:u< x_i< v}\left(Y_i-\hm(x_i)\right)&\geq 0, \label{eq:imp6}\\
\sum_{i:u< x_i< v}x_i\left(Y_i-\hm(x_i)\right)&\geq 0. \label{eq:imp7}
\end{align}
\ee
\el 
\begin{proof}
$(i)$ Let us define $\psi_1(x):=\hm(x)+1$ and $\psi_2(x):=\hm(x)-1$. Plugging $\psi_1(\cdot)$ and $\psi_2(\cdot)$ in Lemma~\ref{CharLemma1}$(iii)$ in place of $\psi(\cdot)$ we can obtain 
 $\sum_{i=1}^n\left(Y_i-\hm(x_i)\right)=0$. Similarly, using $\hm(x)+x$ and $\hm(x)-x$ in place of $\psi(\cdot)$ in Lemma~\ref{CharLemma1}$(iii)$ yields $\sum_{i=1}^n x_i\left(Y_i-\hm(x_i)\right)=0$.
 
 $(ii)$ Let $u$ and $v$ are respectively $x_{k}$ and $x_{l}$. Consider the functions $f_1,f_2$ and $f_3$ plotted in Figure~\ref{Fig0.3}. Note that linear part of $f_2$ on the interval $[x_{k},x_{l}]$ is a section of a line passing through origin.
\begin{figure}[H]
\begin{subfigure}[b]{0.50\textwidth}
 \centering
 {\begin{tikzpicture}
  \draw[->] (-1,0) -- (6,0) node[right] {$x$};
  \draw[->] (0,-1.5) -- (0,2) node[above] {$f_1(x)$};
  \draw[-] (2,-0.05) -- (2,0.05) node[below] {$x_{k}$};
   \draw[] (4,-0.05) -- (4,0.05) node[below] {$x_{l}$};
   \draw[] (1,-0.05) -- (1,0.05) node[below] {$x_{k-1}$};
   \draw[] (5,-0.05) -- (5,0.05) node[below] {$x_{l+1}$};
   \draw[] (0,0.95) -- (0,1.05) node[left] {$1$};
  \draw[domain=-1:1,smooth,variable=\x,blue] plot ({\x},{0});
  \draw[domain=1:2,smooth,variable=\x,blue] plot ({\x},{\x-1});
  \draw[domain=2:4,smooth,variable=\x,blue] plot ({\x},{1});
  \draw[domain=4:5,smooth,variable=\x,blue] plot ({\x},{5-\x});
   \draw[domain=4:6,smooth,variable=\x,blue] plot ({\x},{0});
    \draw[domain=0:2,dashed,variable=\x,blue] plot ({\x},{1}) ; 
    \draw[domain=0:1,dashed,variable=\y,blue] plot ({2},{\y}) ; 
    \draw[domain=0:1,dashed,variable=\y,blue] plot ({4},{\y}) ;
\end{tikzpicture}
\label{Fig0.1}}
\end{subfigure}
\begin{subfigure}[b]{0.50\textwidth}
 \centering
 {\begin{tikzpicture}
  \draw[->] (-1,0) -- (6,0) node[right] {$x$};
  \draw[->] (0,-1.5) -- (0,2.3) node[above] {$f_2(x)$};
  \draw[-] (2,-0.05) -- (2,0.05) node[below] {$x_{k}$};
  \draw[] (4,-0.05) -- (4,0.05) node[below] {$x_{l}$};
   \draw[] (1,-0.05) -- (1,0.05) node[below] {$x_{k-1}$};
   \draw[] (5,-0.05) -- (5,0.05) node[below] {$x_{l+1}$};
   \draw[] (0,1.05) -- (0,2.05) node[below] {$x_{k}$};
  \draw[domain=-1:1,smooth,variable=\x,blue] plot ({\x},{0});
  \draw[domain=1:2,smooth,variable=\x,blue] plot ({\x},{2*\x-2});
  \draw[domain=2:4,smooth,variable=\x,blue] plot ({\x},{\x});
  \draw[domain=4:5,smooth,variable=\x,blue] plot ({\x},{4*(5-\x)});
   \draw[domain=4:6,smooth,variable=\x,blue] plot ({\x},{0});
    \draw[domain=0:2,dashed,variable=\x,blue] plot ({\x},{\x}) ; 
    \draw[domain=0:2,dashed,variable=\y,blue] plot ({2},{\y}) ; 
    \draw[domain=0:4,dashed,variable=\y,blue] plot ({4},{\y}) ;
    \draw[domain=0:2,dashed,variable=\x,blue] plot ({\x},{2}) ;
\end{tikzpicture}
\label{Fig0.2}}
\end{subfigure}
\end{figure}
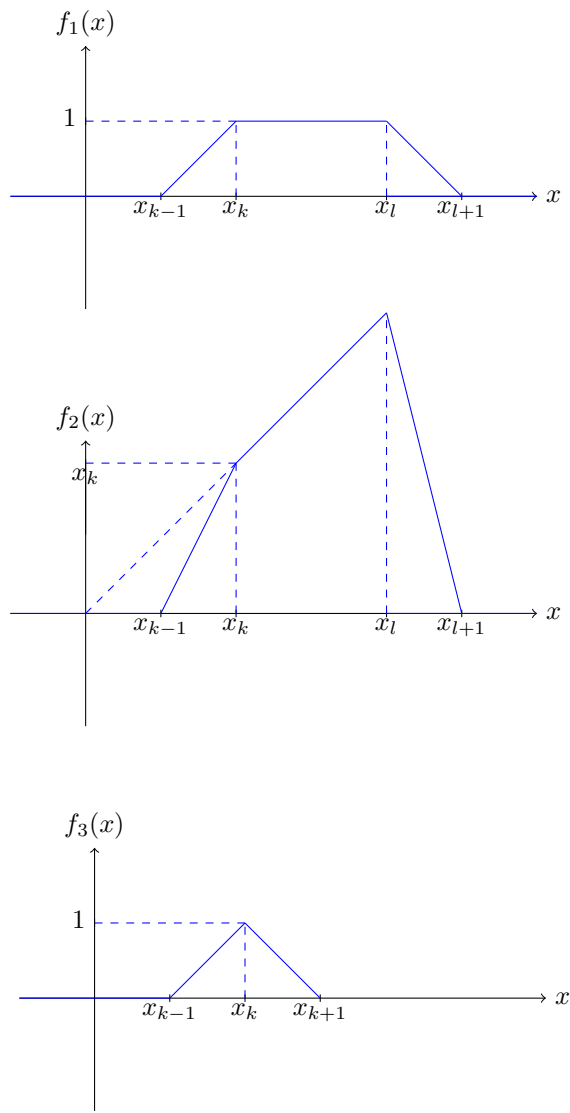
\begin{figure}[H]
\begin{center}
\begin{tikzpicture}
  \draw[->] (-1,0) -- (6,0) node[right] {$x$};
  \draw[->] (0,-1.5) -- (0,2) node[above] {$f_3(x)$};
  \draw[-] (2,-0.05) -- (2,0.05) node[below] {$x_k$};
  \draw[] (1,-0.05) -- (1,0.05) node[below] {$x_{k-1}$};
   \draw[] (3,-0.05) -- (3,0.05) node[below] {$x_{k+1}$};
   \draw[] (0,0.95) -- (0,1.05) node[left] {$1$};
  \draw[domain=-1:1,smooth,variable=\x,blue] plot ({\x},{0});
  \draw[domain=1:2,smooth,variable=\x,blue] plot ({\x},{\x-1});
  \draw[domain=2:3,smooth,variable=\x,blue] plot ({\x},{3-\x});
    \draw[domain=0:2,dashed,variable=\x,blue] plot ({\x},{1}) ;
    \draw[domain=0:1,dashed,variable=\y,blue] plot ({2},{\y}) ;  
\end{tikzpicture}
\caption{Graphs of $f_1$, $f_2$ and $f_3$.}
\label{Fig0.3}
\end{center}
\end{figure}

One can obtain the inequalities in~\eqref{eq:imp1},~\eqref{eq:imp2} and~\eqref{eq:imp5} by replacing $\psi$ with $\hm+\delta f_1, \hm +\delta f_2$ and $\hm+\delta f_3$, respectively, in Lemma~\ref{CharLemma1}$(iii)$ for some sufficiently small $\delta>0$ so that the functions remain convex everywhere inside the interval $[x_1,x_n]$. Similarly, the inequalities in~\eqref{eq:imp6} and~\eqref{eq:imp7} can be derived using  $\hm +\delta g_1$, and $\hm+\delta g_2$, respectively, in Lemma~\ref{CharLemma1}$(iii)$ for some sufficiently small $\delta>0$ where $g_1$ and $g_2$ are plotted in Figure~\ref{Fig0.2}.
\begin{figure}[H]
\begin{subfigure}[b]{0.50\textwidth}
 \centering
 {\begin{tikzpicture}
  \draw[->] (-1,0) -- (6,0) node[right] {$x$};
  \draw[->] (0,-3) -- (0,0.5) node[above] {$g_1(x)$};
  \draw[-] (1,-0.05) -- (1,0.05) node[above] {$x_{k}$};
   \draw[] (5,-0.05) -- (5,0.05) node[above] {$x_{l}$};
   \draw[] (2,-0.05) -- (2,0.05) node[above] {$x_{k+1}$};
   \draw[] (4,-0.05) -- (4,0.05) node[above] {$x_{l-1}$};
   \draw[] (0,-.95) -- (0,-1.05) node[left] {$-1$};
  \draw[domain=-1:1,smooth,variable=\x,blue] plot ({\x},{0});
  \draw[domain=1:2,smooth,variable=\x,blue] plot ({\x},{-\x+1});
  \draw[domain=2:4,smooth,variable=\x,blue] plot ({\x},{-1});
  \draw[domain=4:5,smooth,variable=\x,blue] plot ({\x},{-5+\x});
   \draw[domain=4:6,smooth,variable=\x,blue] plot ({\x},{0});
    \draw[domain=0:2,dashed,variable=\x,blue] plot ({\x},{-1}) ;
    \draw[domain=0:1,dashed,variable=\y,blue] plot ({2},{-\y}) ; 
    \draw[domain=0:1,dashed,variable=\y,blue] plot ({4},{-\y}) ; 
\end{tikzpicture}
\label{Fig0.1}}
\end{subfigure}
\begin{subfigure}[b]{0.50\textwidth}
 \centering
 {\begin{tikzpicture}
  \draw[->] (-1,0) -- (6,0) node[right] {$x$};
  \draw[->] (0,-3) -- (0,.5) node[above] {$g_2(x)$};
  \draw[-] (1,-0.05) -- (1,0.05) node[above] {$x_{k}$};
   \draw[] (5,-0.05) -- (5,0.05) node[above] {$x_{l}$};
   \draw[-] (2,-0.05) -- (2,0.05) node[above] {$x_{k+1}$};
   \draw[] (4,-0.05) -- (4,0.05) node[above] {$x_{l-1}$};
   \draw[] (0,-1.95) -- (0,-2.05) node[above] {$-x_{k+1}$};
  \draw[domain=-1:1,smooth,variable=\x,blue] plot ({\x},{0});
  \draw[domain=1:2,smooth,variable=\x,blue] plot ({\x},{-2*\x+2});
  \draw[domain=2:4,smooth,variable=\x,blue] plot ({\x},{-\x});
  \draw[domain=4:5,smooth,variable=\x,blue] plot ({\x},{-4*(5-\x)});
   \draw[domain=4:6,smooth,variable=\x,blue] plot ({\x},{0});
    \draw[domain=0:2,dashed,variable=\x,blue] plot ({\x},{-\x}) ; 
    \draw[domain=0:2,dashed,variable=\y,blue] plot ({2},{-\y}) ; 
    \draw[domain=0:4,dashed,variable=\y,blue] plot ({4},{-\y}) ;
    \draw[domain=0:2,dashed,variable=\x,blue] plot ({\x},{-2}) ;
\end{tikzpicture}
}
\end{subfigure}
\caption{Graphs of $g_1$ and $g_2$.}
\label{Fig0.2}
\end{figure}
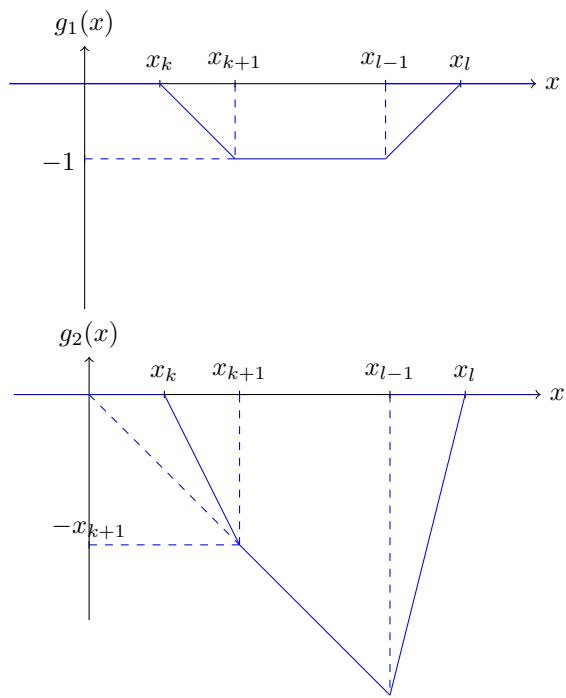
\end{proof}

It is also known that the piecewise affine LSE can be obtained as the projection of $\mathbf{Y}$ onto the closed convex polyhedral cone generated by the functions $\rpm 1$, $\rpm x$, $(x-x_i)_{+}$, where $(x)_{+}:= \max\{x,0\}$ denotes the positive part of $x$; see~\citep{GJW2001b}. The following result gives another representation of the convex LSE. It is a consequence of a more general result stated in~\cite[Proposition~1]{meyer2000} where the underlying polyhedral cone is expressed in terms of linear inequalities; we state the result in terms of the generators of the cone. For completeness, we also give its proof in Section~\ref{pf:Prop2.1}.
 \bp\label{Pr:Representation}
If the set of kinks is found to be $x_{m_1} <\ldots  < x_{m_k}$, $k \ge 0$, then the convex LSE has the 
 following representation:
 \[\hm(x)=\hat{a}+\hat{b}_0x+\sum_{j=1}^k\hat{b}_j(x-x_{m_j})_{+}\]
where $\hat{a}, \hat{b}_0 \in \R$, $\hat{b}_j >0$, for $j=1,\ldots ,k$, are the unconstrained LSEs obtained by minimizing 
\begin{equation}\label{eq:ParaLSE}
\sum_{i=1}^n \left\{Y_i - a  -b_0 x_i  - \sum_{j=1}^k b_j(x_i-x_{m_j})_{+}\right\}^2
\end{equation}
over $a, b_j \in \RR\; (j= 0,\ldots, k)$.
 \ep 
 Although the convex LSE is piecewise affine, the fitted line between two consecutive kink points is not necessarily equal to a simple linear regression fit with the data points in between the two kinks; compare this with isotonic regression where the isotonic LSE is just the average of the response values within the constant `block'.
The following result, proved in Section~\ref{pf:Lem2.3}, shows that the convex LSE is location equivariant under any affine transformation.
\bl\label{equivariance}
Let $\hat{\psi}_n$ be a convex function that minimizes $\sum_{i=1}^n(Y_i + a + b x_i-\psi(x_i))^2$ over $\psi \in \mathcal{K}$, where $a, b \in \R$. Then
   \begin{equation}\label{eq:equiveq}
   \hat{\psi}_n(x_i)=\hm(x_i)+a+bx_i.
   \end{equation}  
  \el


 \section{Local rate of convergence of the LSE}\label{LocalRate}
 In this section we study the rate of convergence of $\hm(x_0)$, where $x_0 \in (0,1)$ is an interior point. For notational convenience, we often use $\mu^\prime$ and $\mu^{\prime\prime}$ to denote first and second derivatives of the convex function $\mu$. In general, we use $\mu^{(r)}$ to denote the $r$-th derivative of $\mu$. 
 By $\hm^{\prime}(x)$, we will always mean $  \hm^{\prime}(x-)$, unless otherwise mentioned.
 We first state the assumption on the errors required for the results in this section:
 \be
 \ii[(\textbf{A1})] $\varepsilon_i$'s are mean zero i.i.d.~random variables with Var$(\varepsilon_i)=\sigma^2 < \infty$. Also, we assume that $\EE[\exp(t\varepsilon_1)]<\infty$, for some $t>0$.
\ee   
Under a uniform fixed design setting, the local rate of convergence of the convex LSE was established in~\citep{MAMMEN1991}. \citep{GJW2001b} generalized the result further and derived the local rate of convergence of the derivative of the LSE. The proof of the following theorem can be found in \citep[Lemma~4.5]{GJW2001b}.    
\bt\label{Ratetheo0}
Suppose that $\mu^{\prime}(x_0)<0$, $\mu^{\prime\prime}(x_0)>0$ and $\mu^{\prime\prime}$ is continuous in a neighborhood of $x_0$. Assume that the design points $x_i=x_{n,i}$ satisfy  \[\frac{c}{n}\leq x_{i,n+1}-x_{i,n}\leq \frac{C}{n},\qquad i=1, \ldots ,n,\]
for some constants $0<c<C<\infty$. Then, under assumption (\textbf{A1}), the LSE $\hm$ satisfies: for each $M>0$,
\begin{align}\label{eq:Conv0}
n^{2/5}\sup_{|t|\leq M}\left|\hm\left(x_0+tn^{-1/5}\right)-\mu(x_0)-tn^{-1/5}\mu^{\prime}(x_0)\right|=O_p(1)
\end{align}
and
\begin{align}\label{eq:Conv00}
n^{1/5}\sup_{|t|\leq M}\left|\hm^{\prime}\left(x_0+tn^{-1/5}\right)-\mu^{\prime}(x_0)\right|=O_p(1).
\end{align} 
\et

\begin{remark}
 It can be noted that one can carry out the proof of Theorem~\ref{Ratetheo0} without assuming $\mu^\prime(x_0)<0$. In the case when $\mu^\prime(x_0)\geq 0$, we just have to add an affine function $-\alpha(x-x_0)$ to the data for a large $\alpha>0$. Using affine equivariance of the convex LSE in Lemma~\ref{equivariance}, we can now say that \eqref{eq:Conv0} and \eqref{eq:Conv00} hold true even when $\hm$ and $\mu$ in Theorem~\ref{Ratetheo0} are replaced by $\hm-\alpha(x-x_0)$ and $\mu-\alpha(x-x_0)$ respectively, thus proving the result for the cases when $\mu^\prime(x_0)\geq 0$.   
\end{remark}
Note that the above result assumes a fixed design setting and only investigates the rate of convergence when $\mu$ has non-vanishing second derivatives. In the following we state the two main results of this section where we assume that the design is random. The first result deals with the case where a certain number of derivatives vanish at $x_0$ (Scenario (a) as mentioned in the Introduction). The second result is applicable when $\mu$ is affine in a neighborhood of $x_0$ (Scenario (b)). 

\bt\label{Ratetheo1} 
Let $x_0 \in (0,1)$ be such that $\mu^{(2)}(x_0)=\ldots =\mu^{(r-1)}(x_0)=0$
 and $\mu^{(r)}(x_0)\neq 0$, $r >2$. Let us assume further that 
 $\mu^{(r)}(x_0)$ is continuous in a neighborhood around $x_0$. Then, under assumption (\textbf{A1}), $\hm(x_0)$ satisfies the following: for each $M>0$,
 \begin{align}\label{eq:Conv1}
n^{\frac{r}{2r+1}}\sup_{|t|\leq M}\left|\hm\left(x_0+tn^{-\frac{1}{2r+1}}\right)-\mu(x_0)\right|=O_p(1)
\end{align}
and
\begin{align}\label{eq:Conv2}
n^{\frac{r-1}{2r+1}}\sup_{|t|\leq M}\left|\hm^{\prime}\left(x_0+tn^{-\frac{1}{2r+1}}\right)-\mu^{\prime}(x_0)\right|=O_p(1).
\end{align} 
\et 
\begin{remark} Let us point out that $r$ should be an even integer and $\mu^{(r)}(x_0)>0$. This is because, by Taylor's theorem, we can say \[\mu^{(2)}(x)= \frac{\mu^{(r)}(x_0)}{(r-2)!}(x-x_0)^{r-2}+o((x-x_0)^{r-1})\] in a some neighborhood of $x_0$. As convexity implies $\mu^{(2)}(x)\geq 0$ for all $x$ whenever $\mu^{(2)}(x)$ exists, $r$ cannot be an odd integer. Further, because of convexity, $\mu^{(r)}(x_0)$ must be greater than $0$.
\end{remark}

\begin{figure}[H]
\centering
\begin{subfigure}{7cm}            
\frame{\includegraphics[width=7cm]{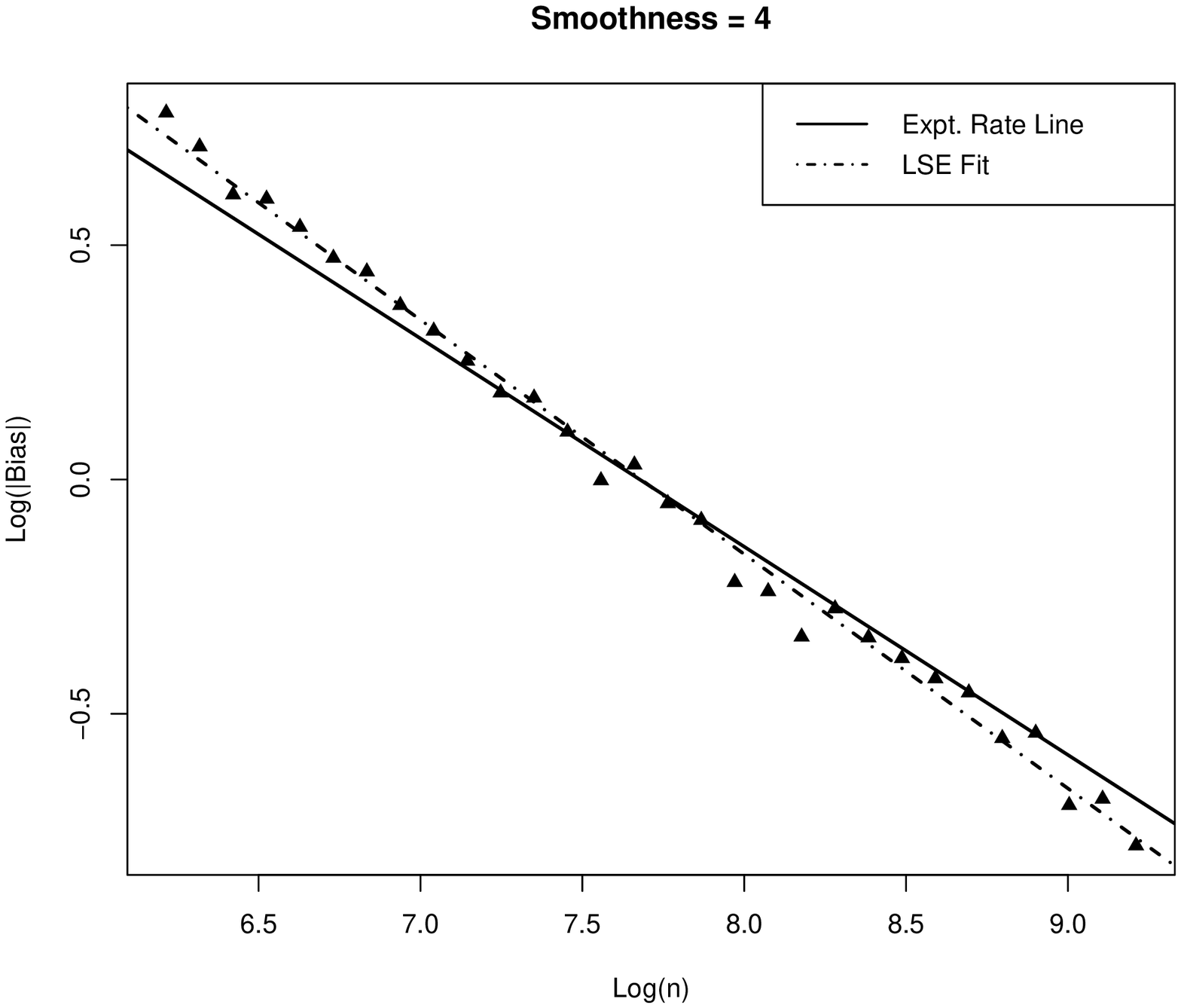}}
\label{Fig:Data1}
\end{subfigure}
\hspace{1cm}
\begin{subfigure}{7cm}
\centering
\frame{\includegraphics[width=7cm]{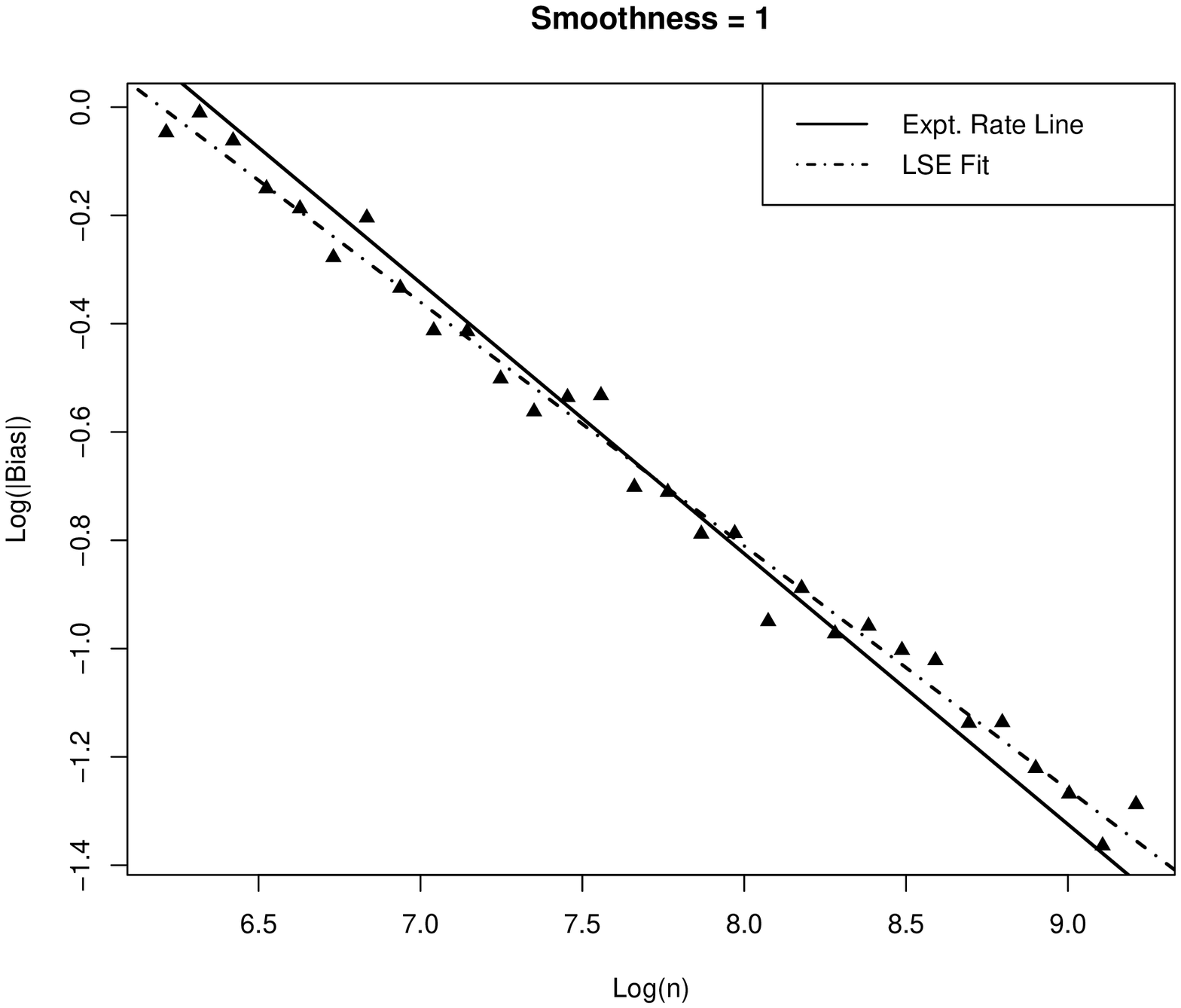}}
\label{Fig:Data2}
\end{subfigure}
\caption{Plots of the logarithms of absolute bias versus the logarithm of sample size. In both scenarios, the slopes of the black solid lines are equal to the negative of the exponents of $n$ in the rates predicted in Theorems~\ref{Ratetheo1} and~\ref{Ratetheo2}.}\label{fig:TOF}
\end{figure}

\bt\label{Ratetheo2} 
Let $\mu$ be affine in the interval $[a,b] \subseteq [0,1]$. Let $x_0 \in (a,b)$ and let $\eta >0$ be such that $[x_0-\eta,x_0+\eta] \subseteq (a,b)$. Then, under assumption (\textbf{A1}), $\hm(x_0)$ satisfies the following: for each $\epsilon>0$ and for each $0<c<\eta$,
 \begin{align}\label{eq:Conv3}
n^{1/2}\sup_{|t|\leq c}\left|\hm\left(x_0+t\right)-\mu(x_0)-\mu^\prime(x_0)t\right|=O_p(1),
\end{align}
and
\begin{align}\label{eq:Conv4}
n^{1/2}\sup_{|t|\leq c}\left|\hm^{\prime}\left(x_0+t\right)-\mu^{\prime}(x_0+t)\right|=O_p(1).
\end{align} 
\et
In Figure~\ref{fig:TOF}, we present two plots illustrating the different rates of convergence of $\hm$ under different assumptions on $\mu$, as indicated in Theorems~\ref{Ratetheo1} and~\ref{Ratetheo2} above. We plot the logarithm of the absolute bias $|\hm\left(x_0\right)-\mu(x_0)|$ in scenarios $(a)$ and $(b)$ with increasing sample size (we take $x_0 = 0.5$). In scenario $(a)$ we choose $r=4$ (indicating `smoothness'  equal to 4) and the model $Y=2(X-0.5)^4+\epsilon$, where $X\sim U(0,1)$ and $\epsilon\sim N(0,1)$ are independent. We draw $200$ replicates for $30$ different sample sizes ranging from $500$ to $10000$. We repeat the same procedure for scenario $(b)$ with the model $Y=2(X-0.5)+\epsilon$ (we represent this by `smoothness' equal to 1). According to Theorem~\ref{Ratetheo1}, one should expect the slope of the best fitted line in the plot with $\mbox{smoothness}=4$ to be equal to $-4/(2 \times 4+1) \approx -0.44$. Similarly, Theorem~\ref{Ratetheo2} predicts that the slope in the second plot should equal $-1/2$. In the simulations the slopes of the best fitted lines came out to be $-.454$ and $-.48$ respectively.       

\subsection{Proofs of the two theorems}
For the sake of clarity, we divide the proofs into several steps. These steps provide the sketches of the proofs and will be closely followed in both the proofs. The proofs of the following lemmas are given in Appendix~\ref{Appendix}.

\noindent \emph{Step} $\mathbf{I}$. Let us define $G:[0,1]\to \RR$ as 
\begin{equation}\label{eq:Theo1}
G(x)=\sum_{i:x_i\leq x}(Y_i-\hat{\mu}_n(x_i))(x-x_i).
\end{equation} 
\bl\label{1stSteplemma}
Let $\Omega$ be the set of all kink points of the convex LSE. Then for all $x\in [x_1,x_n]$, \[G(x)\geq 0,\] and $G(x)=0$ for all $x\in \Omega$. 
\el

\noindent \emph{Step} $\mathbf{II}$.
Let $x_0$ lie in a compact interval $[a,b] \subset (0,1)$. Fix $\epsilon >0$ and let $$T := \Omega \cap [a+\epsilon,b-\epsilon].$$
 \bl\label{2ndSteplemma}
 \begin{equation}\label{eq:Theo2}\underset{x\in T}{\sup}\left|\sum_{i:x_i\leq x} (Y_i-\hat{\mu}_n(x_i) )\right|=O_p\left(\log n\right).
\end{equation}
\el

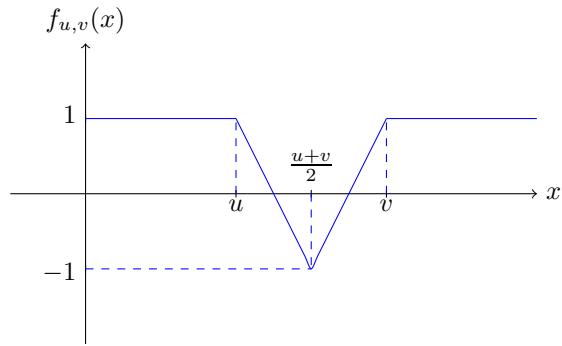
\begin{figure}[H]
\begin{center}
\begin{tikzpicture}
  \draw[->] (-1,0) -- (6,0) node[right] {$x$};
  \draw[->] (0,-2) -- (0,2) node[above] {$f_{u,v}(x)$};
  \draw[-] (2,-0.05) -- (2,0.05) node[below] {$u$};
   \draw[] (4,-0.05) -- (4,0.05) node[below] {$v$};
   \draw[] (0,0.95) -- (0,1.05) node[left] {$1$};
   \draw[] (0,-0.95) -- (0,-1.05) node[left] {$-1$};
   \draw[] (3,-0.05) -- (3,0.05) node[above] {$\frac{u+v}{2}$};
  \draw[domain=0:2,smooth,variable=\x,blue] plot ({\x},{1});
   \draw[domain=2:4,smooth,variable=\x,blue] plot ({\x},{2*abs(\x -3)-1});
    \draw[domain=4:6,smooth,variable=\x,blue] plot ({\x},{1}) ; 
    \draw[domain=0:3,dashed,variable=\x,blue] plot ({\x},{-1}) ;
    \draw[domain=0:1,dashed,variable=\y,blue] plot ({2},{\y}) ;
    \draw[domain=0:1,dashed,variable=\y,blue] plot ({4},{\y}) ;
    \draw[domain=0:1,dashed,variable=\y,blue] plot ({3},{-\y}) ;  
\end{tikzpicture}
\caption{Graph of $f_{u,v}$.}
\label{Fig1}
\end{center}
\end{figure}

\noindent \emph{Step} $\mathbf{III}$. Let us define
\[f_{u,v}(x)=1-\left(2-\dfrac{4}{v-u}\left|x-\dfrac{u+v}{2}\right|\right)_{+}\] and 
\begin{equation}\label{eq:Z_n}
Z(u,v)=n^{-1}\sum_{i=1}^{n}f_{u,v}(x_i)\left(Y_i-\hat{\mu}_n\left(x_i\right)\right).
\end{equation} 
Figure~\ref{Fig1} shows the graph of $f_{u,v}$. 
\bl\label{3rdSteplemma}
We have the following results:
\bei
\ii $Z(u,v)\leq 0$, for $u<v\in T$.
\ii Define $$U_n:=\sup\{x\in \Omega:x\leq x_0\} \qquad \mbox{and} \qquad V_n:=\inf\{x\in \Omega:x>x_0\}.$$ If $\{U_n\}$, $\{V_n\}$ satisfy \begin{equation}\label{eq:V_n-U_n}
V_n-U_n=O_p(c_n), \quad\text{and}\quad \frac{1}{V_n-U_n}=O_p(c_n^{-1})
\end{equation} 
for some bounded sequence of positive real
 numbers $\{c_n\}_{n=1}^\infty$, then \[\left|Z(U_n,V_n)- n^{-1}\sum_{U_n\leq x_i\leq V_n}f_{U_n,V_n}
 (x_i)Y_i\right|=\sqrt{V_n-U_n}O_p ({n}^{-1/2})+O_p({n}^{-1}{\log n}).\]
\ee
\el
The above lemma is crucial in the sequel. In some sense, the above display helps us ``localize" to a neighborhood of $x_0$. From now on we will mainly study the localized random variable $n^{-1}\sum_{U_n\leq x_i\leq V_n}f_{U_n,V_n} (x_i)Y_i$ and use the fact that $Z(U_n,V_n)\leq 0$ to derive our result. 

\noindent  \emph{Step} $\mathbf{IV}$. We expand $n^{-1}\sum_{U_n\leq
  x_i\leq V_n}f_{U_n,V_n}(x_i)Y_i$ into two parts, namely $Z_1(U_n,V_n)$ and $Z_2(U_n,V_n)$, such that
 \begin{equation}\label{eq:expanRate}
  Z(U_n,V_n)=Z_1(U_n,V_n)+Z_2(U_n,V_n)+\sqrt{V_n-U_n}O_p ({n}^{-1/2})+O_p({n}^{-1}{\log n})
 \end{equation}
 where 
 \begin{equation}\label{eq:expanRate1}
 Z_1(U_n,V_n)=\frac{1}{n}\sum_{U_n\leq x_i\leq V_n}f_{U_n,V_n}(x_i)\mu (x_i)\end{equation}
 and 
 \begin{equation}\label{eq:expanRate2}
 Z_2(U_n,V_n)=\frac{1}{n}\sum_{U_n\leq x_i\leq V_n}f_{U_n,V_n}(x_i)\varepsilon_i.
 \end{equation}

\noindent  \emph{Step} $\mathbf{V}$. Utilizing the inequality $Z(U_n,V_n)\leq 0$ we will find out the rate at which the two consecutive kink points around $x_0$ come close to each other.

\noindent \emph{Step} $\mathbf{VI}$. Again utilizing the result in \emph{Step} $\mathbf{I}$, we will find the rate at which 
 \[\underset{\mathcal{U}_n\leq x\leq \mathcal{V}_n}{\inf}\left|\hm(x)-\mu(x)\right|\]
 will converge to zero when two sequence of kinks $\{\mathcal{U}_n\}\subset \{x\in\Omega:x\le x_0\}$ and $\{\mathcal{V}_n\}\subset\{x\in\Omega:x>x_0\}$ approach each other at a certain rate.
 \vspace{0.2cm}

\noindent \emph{Step} $\mathbf{VII}$. Finally, we will derive the rate of convergence of the derivative $\hm^\prime(x_0)$ and will utilize this to find the rate of convergence of $\hm(x_0)$.


 \subsubsection{Proof of the Theorem \ref{Ratetheo1}}
Without loss of generality, we can also assume $\mu^\prime(x_0)=0$ thanks to the affine equivariance property of convex LSE proved in Lemma~\ref{equivariance}. We will first show that $V_n - U_n \overset{a.s.}{\rightarrow}0$. Suppose that the true convex function has a change in slope in an open interval around $x_0$. Then, almost surely (a.s.) there will exist a bend point of $\hm$ in that interval, for sufficiently large $n$, as $\hm$ converges uniformly to $\mu$ on compact sets contained in the interior of $(0,1)$; see \citep{SS11} and \citep[Lemma 5]{MAMMEN1991}. 

Since $\mu^{(r)}(\cdot)$ is continuous at $x_0$, there exists $\delta >0$ such that for all $x\in(x_0-\delta,x_0+\delta)$, $\mu^{(r)}(x)>\frac{1}{2}\mu^{(r)}(x_0)$. As a result, using an $(r-1)$-fold integral, $$\mu'(x_0 - \delta/3) - \mu'(x_0 - \delta) \ge c \mu^{(r)}(x_0)\delta^{r-1}$$ for some constant $c >0$. Similarly, the change in slope in the interval $(x_0+\delta/3,x_0+\delta)$ has the same lower bound. Hence for sufficiently large $n$,  with probability one, there exists at least two bend points around $x_0$ within a distance less than $2\delta$. In fact, the above observation holds for any $0<\epsilon<\delta$. So for all $\epsilon<\delta$ we can argue that $\PP(V_n-U_n>\epsilon\mbox{ i.o.})= 0$. In particular, the union of all such events $\{V_n-U_n>\epsilon \mbox{ i.o.}\}$ will have zero probability whenever $\epsilon$ varies over set of all rationals. Hence, $V_n-U_n \overset{a.s.}{\rightarrow}0$.    

By using a Taylor series expansion of $\mu$ in \eqref{eq:expanRate1} and the continuity of $\mu^{(r)}$ around $x_0$, we get
 \begin{align*}
 Z_1\left(U_n,V_n\right)&=n^{-1}\sum_{U_n\leq x_i\leq V_n}f_{U_n,V_n}(x_i)\left(\frac{1}{r!}\mu^{ 
 \left(r\right)}(x_0)\left(x_i-U_n\right)^r+o\left((V_n-U_n)^{r}\right)\right).
 \end{align*}
For $\delta >0$, consider the class of functions 
 \[\mathfrak{F}_{\delta}=\{f_{u,v}(x) \mathbf{1}_{[u,v]}(x)(x-u)^{r}:\p x_0-\delta<u\leq x_0\leq v<x_0+\delta\}.\]
An envelope function for $\mathfrak{F}_{\delta}$ can be taken as 
 \[F_{\delta}(x) = (2\delta)^{r} \mathbf{1}_{[x_0-\delta,x_0+\delta]}(x),\] 
 so that 
 \[ \EE\left[F_{\delta}^2(X)\right]=(2\delta)^{2r}\{F(x_0+\delta)-F(x_0-\delta)\} = 2^{2r} \delta^{2r+1}.\]
Using Theorem 2.14.1 of~\citep{VAN&WELL2000}, we have
\begin{align}\label{eq:Impone}
\EE\Big[\Big(\underset{g\in \mathfrak{F}_{\delta}}{\sup}\left|\left(\mathbb{P}_n-P\right)g\right|\Big)^2\Big]\leq \frac{K}{n} \EE\left[F_{\delta}^2(X)\right] =O(n^{-1}\delta^{2r+1}),
 \end{align}
 where $K >0$ is a constant.
 \bp\label{Prop:1st}
There exists $\delta_0>0$ such that for each $\epsilon>0$, there exists random variables $\{M_n\}$ of the order $O_p(1)$ such that the following holds  for all $u,v$, where $x_0-\delta_0<u\leq x_0\leq v<x_0+\delta_0$:
\begin{equation}\label{eq:Pn-P}
\left|\left(\mathbb{P}_n - P\right)f_{u,v}(X)\mathbf{1}_{[u,v]}(X)(X-u)^r\right|\leq \epsilon(v-u)^{r+1}+n^{-(r+1)}M_n^2.
\end{equation}
\ep
Now, we can show that 
\[P [f_{u,v}(X)\mathbf{1}_{[u,v]}(X)(X-u)^r]= c \mu^{(r)}(x_0)(v-u)^{r+1}\]
as $X \stackrel{}{\sim}$ Uniform$(0,1)$. Using Proposition~\ref{Prop:1st}, we can write
 \begin{equation}\label{eq:Z_1rate}
 Z_1\left(U_n,V_n\right)=\frac{1}{2^{r+1}(r+1)!}\mu^{(r)}(x_0)(V_n-U_n)^{r+1}+\epsilon(V_n-U_n)^{r+1}+O_p\left(
 n^{-(r+1)}\right).
 \end{equation}

\bl\label{Lem:1stResult1} We have
\begin{equation}\label{eq:kinkRate1}
  V_n-U_n=O_p(n^{-\frac{1}{2r+1}}).
\end{equation} 
\el 
From Lemma \ref{1stSteplemma} it is clear that $G(x)=0$ for all $x\in T$. In particular, $G(U_n)=0$ and
 $G(V_n)=0$ and as a result $G(V_n)-G(U_n)=0$. The following result is similar in spirit to~\citep[Lemma 4.3]{GJW2001b}.
\bl\label{Lem:1stResult2} 
Let $\mathcal{U}_n$ and $\mathcal{V}_n$ be two sequences of kinks, for $n \ge 1$, such that $\mathcal{V}_n>\mathcal{U}_n$ and $cn^{-1/(2r+1)}<\mathcal{V}_n- \mathcal{U}_n=O_p(n^{-1/(2r+1)})$ for some $c>0$. Then
  \begin{equation}\label{eq:infRate1}\underset{\mathcal{U}_n\leq x\leq \mathcal{V}_n}{\inf}\left|\hat{\mu}_n(x)-\mu(x)\right|=O_p\left(n^{-\frac{r}{2r+1}}\right).
  \end{equation}
  \el

\begin{figure}[H]
\begin{center}
\begin{tikzpicture}
  \draw[-] (-8,0) -- (8,0);
  \draw[-] (0,-0.05) -- (0,0.05) node[below] {$x_0$};
   \draw[] (-2,-0.05) -- (-2,0.05) node[below] {$\sigma_{n,-1}$};
   \draw[decorate,decoration={brace,amplitude=10pt}] (-1.7,0.1) -- (0,0.1) node[above] {$Ml_n$};
   \draw[] (2,-0.05) -- (2,0.05) node[below] {$\sigma_{n,1}$};
   \draw[decorate,decoration={brace,amplitude=10pt}] (0,0.1) -- (1.7,0.1) node[above] {$Ml_n$};
   \draw[] (-3.5,-0.05) -- (-3.5,0.05) node[below] {$\sigma_{n,-2}$};
   \draw[] (3.5,-0.05) -- (3.5,0.05) node[below] {$\sigma_{n,2}$};
   \draw[] (-5.5,-0.05) -- (-5.5,0.05) node[below] {$\sigma_{n,-3}$};
   \draw[decorate,decoration={brace,amplitude=10pt}] (-4.7,0.1) -- (-3.5,0.1) node[above] {$l_n$};
   \draw[] (5.5,-0.05) -- (5.5,0.05) node[below] {$\sigma_{n,3}$};
   \draw[decorate,decoration={brace,amplitude=10pt}] (3.5,0.1) -- (4.7,0.1) node[above] {$l_n$};
   \draw[] (-7,-0.05) -- (-7,0.05) node[below] {$\sigma_{n,-4}$};
   \draw[] (7,-0.05) -- (7,0.05) node[below] {$\sigma_{n,4}$};
\end{tikzpicture}
\caption{Diagram showing the positions of the kink points. Here $l_n:=n^{-1/(2r+1)}$.}
\label{Fig2}
\end{center}
\end{figure}
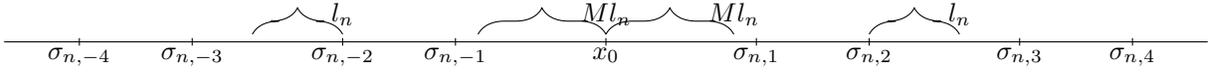

We will now derive the rate of convergence of $\hm^\prime$ around a neighborhood of $x_0$. Fix $M>0$. Let us denote the left nearest kink point to $x_0-Mn^{-1/(2r+1)}$ and the right nearest kink point to $x_0+Mn^{-1/(2r+1)}$ by $\sigma_{n,-1}$ and $\sigma_{n,1}$ respectively. Also, let us denote the left nearest kink point to $\sigma_{n,-1}$ by $\sigma_{n,-2}$ and the right nearest kink point to $\sigma_{n,1}$ by $\sigma_{n,2}$. Further, let us name the nearest kink on the right side of $\sigma_{n,2}+n^{- {1}/{(2r+1)}}$ by $\sigma_{n,3}$, and the nearest kink to the left of $\sigma_{n,-2}-n^{-{1}/{(2r+1)}}$ by $\sigma_{n,-3}$. Further, denote the left nearest kink to $\sigma_{n,-3}$ by $\sigma_{n,-4}$ and the right nearest kink to $\sigma_{n,3}$ by $\sigma_{n,4}$. Figure~\ref{Fig2} explains the above notation pictorially.
  
 Note that all the results proved till now for $U_n=\sup\{x\le x_0:\hm^{\prime}(x-)<\hm^{\prime}(x+)\}$ and $V_n=\inf\{x\geq x_0:\hm^{\prime}(x-)<
 \hm^{\prime}(x+)\}$ also can be proved if we take $U_n$ as $\sup\{x\le \xi_n:\hm^{\prime}(x-)<\hm^{\prime}(x+)\}$ and $V_n$ as $\inf\{x\geq \xi_n:
 \hm^{\prime}(x-)<\hm^{\prime}(x+)\}$ for some sequence $\{\xi_n\}^\infty_{n=1}$ such that $\xi_n\stackrel{}{\longrightarrow}x_0$ as $n\to\infty$. Hence from 
 Lemma~\ref{Lem:1stResult2}, it is clear that $\sigma_{n,4}-\sigma_{n,-4}=O_p(n^{-\frac{1}{2r+1}})$.
We will name the point of minimum of $|\hat{\mu}_{n}(x)-\mu(x)|$ in the interval $[\sigma_{n,i},\sigma_{n,i+1}]$ as $\eta_{n,i+1}$. For any $t\in \RR$, such that $|t|\leq M$, 
 \begin{eqnarray*}
 \hat{\mu}^{\prime}_{n}(x_0+tn^{-\frac{1}{2r+1}})&\geq & \frac{\hat{\mu}_n(\eta_{n,-1})-\hat{\mu}_n(\eta_{n,-3})}{\eta_{n,-1}-\eta_{n,-3}}\\
 &\geq & \frac{\mu(\eta_{n,-1})-\mu(\eta_{n,-3})}{\eta_{n,-1}-\eta_{n,-3}}+\frac{\hat{\mu}_n(\eta_{n,-1})-\mu(\eta_{n,-1})-\hat{\mu}_n(\eta_{n,-3})+\mu(\eta_{n,-3})}{\eta_{n,-1}-\eta_{n,-3}}\\
 &\geq & \mu^{\prime}(x_0)-c_\epsilon n^{-\frac{r-1}{2r+1}}
 \end{eqnarray*}
for some $c_\epsilon>0$ with probability greater than $1-\epsilon$.
 Similarly using $\eta_{n,2}$ and $\eta_{n,4}$, it is easy to see that 
 \[\hat{\mu}^{\prime}_{n}(x_0+tn^{-\frac{1}{2r+1}})\leq \mu^{\prime}(x_0)+c_\epsilon n^{-\frac{r-1}{2r+1}}\]
 holds  with probability greater than $1-\epsilon$.
 Hence we get
 \begin{equation}\label{eq:derirate}\sup_{|t|\le M}|\hat{\mu}^{\prime}_n(x_0+tn^{-\frac{1}{2r+1}})-\mu^\prime(x_0)|=O_p(n^{-\frac{r-1}{2r+1}}).
 \end{equation}
 For the pointwise rate of convergence of the convex function, we will make use of the last two results.
 We will only prove that given any $\epsilon>0$, there exists $K_\epsilon>0$ such that $\hat{\mu}_n(x_0)\geq \mu(x_0)-K_\epsilon n^{-\frac{r}{2r+1}}$ for some finite $K_\epsilon$ (the other side can be proved similarly (see \citep[Lemma 4.4]{GJW2001b})):
 \begin{eqnarray*}
 \hat{\mu}_n(x_0)&\geq & \hat{\mu}_n(\eta_{n,-1})+\hat{\mu}^{\prime}_n(\eta_{n,-1})(x_0-\eta_{n,-1})\\
 &\geq & \left(\hat{\mu}_n(\eta_{n,-1})-\mu(\eta_{n,-1})\right)+\mu(\eta_{n,-1})+\hat{\mu}^{\prime}_n(\eta_{n,-1}) (x_0-\eta_{n,-1})\\
 &\geq &\mu(x_0)-K_\epsilon n^{-\frac{r}{2r+1}}.
 \end{eqnarray*}
The last inequality follows from Lemma~\ref{Lem:1stResult2} and~\eqref{eq:derirate} as it is clear from the definition of $\eta_{n,-1}$ that $Mn^{-1/(2r+1)}<|x_0-\eta_{n,-1}|=O_p(n^{-1/(2r+1)})$.

\subsubsection{Proof of Theorem~\ref{Ratetheo2}}
Let us assume that $x_0 \in (a,b) \subseteq [0,1]$ and that $\mu$ is  affine on $[a,b]$. If $0<a<b<1$, we will assume that $\mu$ has a strict change of slope at $a$ and $b$, i.e., $[a,b]$ is the maximal interval around $x_0$ on which $\mu$ is affine. Let $\mathcal{U}^{1}_n$ and $\mathcal{V}^{1}_n$ be the left nearest and the right nearest kink points to $a$ respectively. If $a$ happens to be equal to $0$, then $\mathcal{U}^{1}_n$ and $\mathcal{V}^{1}_n$ are both defined as the nearest kink point to $0$. In case there is no kink point on the left side (right side) of $a$, we will let $\mathcal{U}^{1}_n$ ($\mathcal{V}^{1}_n$) be equal to $a$. Let $\mathcal{U}^{2}_n$ and $\mathcal{V}^{2}_n$ be defined similarly for $b$. Let 
$a<x_0-\eta<x_0+\eta<b$ for some $\eta>0$. 

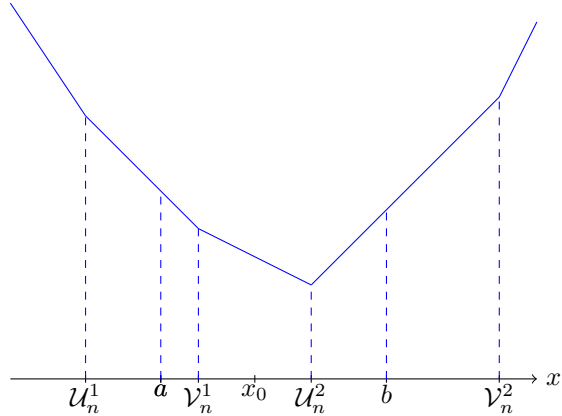
\begin{figure}[H]
\begin{center}
\begin{tikzpicture}
  \draw[->] (-1,0) -- (6,0) node[right] {$x$};
  \draw[-] (0,-0.05) -- (0,0.05) node[below] {$\mathcal{U}^{1}_n$};
  \draw[-] (1.5,-0.05) -- (1.5,0.05) node[below] {$\mathcal{V}^{1}_n$};
  \draw[-] (1,-0.05) -- (1,0.05) node[below] {$a$};
  \draw[-] (1,-0.05) -- (1,0.05) node[below] {$a$};
  \draw[-] (2.25,-0.05) -- (2.25,0.05) node[below] {$x_0$};
   \draw[-] (4,-0.05) -- (4,0.05) node[below] {$b$};
   \draw[-] (3,-0.05) -- (3,0.05) node[below] {$\mathcal{U}^{2}_n$};
   \draw[-] (5.5,-0.05) -- (5.5,0.05) node[below] {$\mathcal{V}^{2}_n$};
  \draw[domain=1.5:3,smooth,variable=\x,blue] plot ({\x},{-0.5*\x+2.75});
  \draw[domain=0:1.5,smooth,variable=\x,blue] plot ({\x},{-\x+3.5});
  \draw[domain=-1:0,smooth,variable=\x,blue] plot ({\x},{-1.5*\x+3.5});
    \draw[domain=3:5.5,smooth,variable=\x,blue] plot ({\x},{\x-1.75}) ;
    \draw[domain=5.5:6,smooth,variable=\x,blue] plot ({\x},{2*\x-7.25}) ;
    \draw[domain=0:3.5,dashed,variable=\y,blue] plot ({0},{\y}) ;
    \draw[domain=0:2.5,dashed,variable=\y,blue] plot ({1},{\y}) ;
    \draw[domain=0:2,dashed,variable=\y,blue] plot ({1.5},{\y}) ;
    \draw[domain=0:1.25,dashed,variable=\y,blue] plot ({3},{\y}) ;  
    \draw[domain=0:2.25,dashed,variable=\y,blue] plot ({4},{\y}) ;
    \draw[domain=0:3.75,dashed,variable=\y,blue] plot ({5.5},{\y}) ;
\end{tikzpicture}
\caption{Graph of one possible instance of $\hm$ around $x_0$. Note that in this picture $\mathcal{U}_n=\mathcal{V}^{1}_n$ and $\mathcal{V}_n=\mathcal{U}^{2}_n$.}
\label{Fig3}
\end{center}
\end{figure}

 Due to the convexity and strict change of slope of $\mu$ at $a$ and $b$, whenever $0<a<b<1$, and the a.s.~consistency of $\hat \mu_n$, it is clear that $\mathcal{U}_n:=\big[\arg \min_{\{\mathcal{U}^{1}_n,\mathcal{V}^{1}_n\}}|x-a|\big]\vee a$ and $\mathcal{V}_n:=\big[\arg \min_{\{\mathcal{U}^{2}_n,\mathcal{V}^{2}_n\}}|x-b|\big]\wedge b$ converge a.s.~to $a$ and $b$ respectively. If $a=0$ or $b=1$, it may happen that both $\mathcal{U}_n$ and $\mathcal{V}_n$ defined above end up being the same point. One can also prove the a.s.~convergence of $\mathcal{U}_n$ and $\mathcal{V}_n$ to $a$ and $b$ accordingly whenever $a=0$ or $b=1$, if we redefine $\mathcal{V}_n:=b$ when it is closer to $a$ relative to $b$ and $\mathcal{U}_n:=a$ for the other case. Figure~\ref{Fig3} explains the above notation pictorially.
 
 Define $$T := (\Omega\cap [a,b])\cup\{\mathcal{U}_n,\mathcal{V}_n\}$$ and fix $\epsilon>0$ so that $[x_0-\eta,x_0+\eta]\subset (a+\epsilon,b-\epsilon)$. Let $\{\xi_n\}^\infty_{n=1}$ be any sequence in the interval
 $[a+\epsilon, b-\epsilon]$. Let us define $$U_n:=\sup\{x\le \xi_n:x\in T\} \qquad \mbox{ and } \qquad V_n:=\inf\{x> \xi_n:x\in T\}.$$
Using similar arguments as that in the proof of Lemma~\ref{3rdSteplemma}, one can show that
 \begin{equation}\label{eq:linear1steq}
 \left|Z(U_n,V_n)-n^{-1}\sum_{U_n\leq x_i\leq V_n}f_{U_n,V_n}(x_i)Y_i\right|=O_p\left(\sqrt
 {\frac{V_n-U_n}{n}}\right)+O_p\left(\frac{\log n}{n}\right)
 \end{equation}
 where $Z(\cdot,\cdot)$ is defined in~\eqref{eq:Z_n}. Under the assumption of linearity of $\mu$ inside the interval $[a,b]$, using the same techniques as in the proof of Lemma~\ref{3rdSteplemma}$(ii)$ it can be shown that
 \[Z_1(U_n,V_n)=\left\{\sqrt{(V_n-U_n)}+(V_n-U_n)^{3/2}\right\}O_p\left(\frac{1}{\sqrt{n}}\right).\]
As $Z(U_n,V_n)\leq 0$, we can compare the order of $Z_2(U_n,V_n)$ (see \ref{eq:Z_2rate}) with other terms in the expansion of $Z(U_n,V_n)$ in~\eqref{eq:expanRate}. Ignoring the smaller order terms (a more elaborate analysis can be done to tackle the smaller order terms) leads to the following inequality:
 \[\sqrt{\frac{V_n-U_n}{n}}\lesssim \frac{(V_n-U_n)^{3/2}}{\sqrt{n}}\]
 which yields
 \begin{equation}\label{eq:V_n-U_n}
 \frac{1}{V_n-U_n}=O_p(1).
 \end{equation} 
 Now let us choose a random sequence $\{\xi_n\}^\infty_{n=1}$ of real numbers such that $\xi_n\in [u_n,v_n]$ where $[u_n, v_n]$ is the interval of smallest length for all consecutive $u, v\in T\cap[a+\epsilon, b-\epsilon]$.  Repeating all the arguments for $\{U_n\}^\infty_{n=1}$ and $\{V_n\}^\infty_{n=1}$ defined through $\{\xi_n\}^\infty_{n=1}$ it follows that~\eqref{eq:V_n-U_n} holds, and as a consequence we have
 \begin{equation}\label{eq:linear2ndeq}\frac{1}{\underset{\substack{U_n\neq V_n \in T\cap(a+\epsilon,b-\epsilon)}}{\inf}\left(V_n-U_n\right)}=O_p(1).
 \end{equation} 
Note that \eqref{eq:linear2ndeq} implies that the  length of each of the linear sections of $\hm$ in the interval $(a+\epsilon,b-\epsilon)$ is $O_p(1)$. Thus, the least squares regression lines fitted on each of these intervals will be $\sqrt{n}$-consistent, converging to $\mu$. On top of that, if one can obtain tight bounds on the deviation of these least squares regression lines from $\hm$ on each of these affine sections of $\hm$, then it would be possible to derive the rate of convergence of $\hm$ to $\mu$ at $x_0$. In the subsequent discussion, we try to make this intuition rigorous. 
Let us consider two end points $u$ and $v$ of any affine part of $\hm$. Note that \eqref{eq:imp1} and \eqref{eq:imp2} in conjunction with \eqref{eq:imp6} and \eqref{eq:imp7} in Lemma~\ref{Char3} imply  
\begin{align}
\left|\sum_{i:u\leq x_i\leq v}\left(Y_i-\hm(x_i)\right)\right|&\leq  \left|Y_{k_1}-\hm(x_{k_1})\right|+\left|Y_{k_2}-\hm(x_{k_2})\right|,\label{eq:imp8}\\
\left|\sum_{i:u\leq x_i\leq v}x_i\left(Y_i-\hm(x_i)\right)\right|&\leq |x_{k_1}|\left|Y_{k_1}-\hm(x_{k_1})\right|+|x_{k_2}|\left|Y_{k_2}-\hm(x_{k_2})\right|, \label{eq:imp9}
\end{align}
where $k_1$ and $k_2$ denote the indices of $u$ and $v$ respectively.
\bd\label{Shortdef}
Define  $\t$ and $\tt$ in the following way:
\begin{align*}
\t & :=\sum_{i:u\leq x_i\leq v}\left(Y_i-\hm(x_i)\right)\\
\tt &:=\sum_{i:u\leq x_i\leq v}x_i\left(Y_i-\hm(x_i)\right).
\end{align*}
\ed
We will denote the mean of the $x_i$'s in the interval $[u,v]$ by $\bar{x}$, and by $\hat{a}_{ls}$ \& $\hat{b}_{ls}$ the simple linear LSEs of the intercept and the slope parameters fitted over the data points in the interval $[u,v]$. 
\bl\label{difflemma} 
We have
\[\sup_{x\in [u,v]}\left|\hat{a}_{ls}+\hat{b}_{ls}x-\hm(x)\right|\leq |v-u|\left|\frac{\bar{x}\t-\tt}{\sum_{u\leq x_i\leq v}(x_i-\bar{x})^2}\right|+\frac{\left|\t\right|}{k_2-k_1+1}.\] 
\el

\bl\label{Hingelemma}
Let $u,v$ be two consecutive kink points of $\hm$ and let $\hat{a}_{ls}+\hat{b}_{ls}x$ be the least squares regression line fitted over the data points in the interval $[u,v]$. Further assume that $1/(v-u)=O_p(1)$. Then, 
\begin{equation}\label{eq:linear6theq}\sup_{x\in [u,v]}\left|\hat{a}_{ls}+\hat{b}_{ls}x-\hm(x)\right|=O_p\left(\frac{\log n}{n}\right).
\end{equation}

\el
Next we apply Lemma~\ref{difflemma} and Lemma~\ref{Hingelemma} to complete the proof. It is clear from \eqref{eq:linear2ndeq} that for any given $\gamma \in (0,1)$, for sufficiently 
 large $n$, there exists $m_\gamma>0$ such that \begin{align}\label{eq:linear7theq}
 [u_n,u_n+m_\gamma]\subset [u_n,v_n]\subset [\mathcal{U}_n, \mathcal{V}_n]\subseteq [a,b]
 \end{align}
 holds with probability greater than $1-\gamma$ where $u_n$, $v_n$ are defined immediately after~\eqref{eq:V_n-U_n} and $\mathcal{U}_n$, $\mathcal{V}_n$ are specified at the beginning this subsection. Note that \eqref{eq:linear7theq} implies the length of each of the affine sections of $\hm$ in the interval $[x_0-\eta,x_0+\eta]$ is at least $m_\gamma$ with probability greater than $1-\gamma$. So $O_p(n)$ realization of $x_i$'s  fall inside of each of these intervals. Hence, the LSEs $\hat{a}_{ls}$ and $\hat{b}_{ls}$ on any of these affine sections of $\hm$ would be $\sqrt{n}$-consistent. 
 
Let $\hat{a}^{(t)}_{ls}$ and $\hat{b}^{(t)}_{ls}$ be the LSEs of the intercept and the slope parameters for the simple linear regression model fitted over the data points in the interval $[U^{(t)}_n,V^{(t)}_n]$, where $U^{(t)}_n:=\sup\{x\le x_0 + t:x\in T\}$ and $V^{(t)}_n:=\inf\{x> x_0+t:x\in T\}$. Then, for all large $n$,
 \begin{eqnarray}\label{eq:finaleqratetheo2}
  &&\PP \left(\underset{|t|\leq \eta}{\sup}\sqrt{n}|\hm(x_0+t)-\mu(x_0)-\mu^\prime(x_0)t|>M\right)\nonumber\\
&\leq & \PP\left(\underset{|t|\leq \eta}{\sup}\sqrt{n}|\hat{a}^{(t)}_{ls}+\hat{b}^{(t)}_{ls}(x_0+t)-\mu(x_0)-\mu^\prime(x_0)t|>M/2\right)\nonumber\\&&+\; \PP\left(\underset{|t|\leq \eta}{\sup}\sqrt{n}|\hat{a}^{(t)}_{ls}+\hat{b}^{(t)}_{ls}(x_0+t)-\hm(x_0)|>M/2\right)\nonumber\\
 &\leq & \PP\left(\underset{|t|\leq \eta}{\sup}\sqrt{n}|\hat{a}^{(t)}_{ls}+\hat{b}^{(t)}_{ls}(x_0+t)-\mu(x_0)-\mu^\prime(x_0)t|>M/2\right)+2\gamma,
\end{eqnarray}
where the last inequality follows from \eqref{eq:linear6theq} and the fact that $v_n-u_n\geq m_\gamma$ with probability $1-\gamma$ for all large $n$.
  Note that the first term in the right side of \eqref{eq:finaleqratetheo2} can be made arbitrarily small by choosing $M$ sufficiently large. Hence, 
  \[\underset{|t|\leq \eta}{\sup}\sqrt{n}|\hm(x_0+t)-\mu(x_0)-\mu^\prime(x_0)t|=O_p(1).\]

\section{Asymptotic distributions}\label{AsympDist}
In this section we will establish the pointwise asymptotic theory of the estimators in both the scenarios (a) and (b), as mentioned in the Introduction. The proof of the main result in this section is divided into three steps, similar to that of the proof of the pointwise distribution theory in~\citep{GJW2001b}. We first  define localized processes whose double and third derivatives at zero arise as the asymptotic limits of the
properly scaled (and centered) LSE and its derivative. 
\bt\label{AsymTheo1}
Let $X_{(r)}(t)=W(t)+(r+2)t^{r+1}$, for $ t \in \R$, where $W(t)$ is standard two-sided Brownian motion starting from 0, and let $Y_{(r)}$ be the integral of $X_{(r)}$, satisfying $Y_{(r)}(0)=0$, i.e., $Y_{(r)}(t)=\int_0^t W(t) dt +t^{r+2}$ for $t \in \R$. Then there exists an a.s.~uniquely defined random continuous function $H_{(r)}$ satisfying the following conditions:
\be
\ii  $H_{(r)}$ is everywhere above the function $Y$; i.e.,
\[H_{(r)}(t)\geq Y_{(r)}(t) \quad\text{for all $t\in\RR$},\]
\ii  $H_{(r)}$ has a convex second derivative, and with probability 1, $H_{(r)}$ is three times differentiable at $t=0$,
\ii  $H_{(r)}$ satisfies 
\[\int_{\RR}\left\{H_{(r)}(t)-Y_{(r)}(t)\right\}dH_{(r)}^{\prime\prime\prime}(t)=0.\]
\ee
\et
Following~\citep{GJW2001b} we will call $H_{(r)}$ to be the `invelope' process of $Y_{(r)}$.

\begin{remark}\label{AsymTheo2remark}
Note that the `invelope' process of $Y_{(r)}$ defined in Theorem \ref{AsymTheo1} is just the analogue of the 'invelope' for the process $Y \equiv Y_{(2)}$ defined in~\citep{GJW2001b}. The proofs of the existence and uniqueness of $H_{(r)}$ will follow the exact same steps involved in proving the analogous results for $H \equiv H_{(2)}$ in~\citep{GJW2001a}. Completely rigorous proof of Theorem~\ref{AsymTheo1} is beyond the scope of this paper. Hence, we omit the proof of this result and assume that the result holds for the rest of the paper.   
\end{remark}
\bt[Asymptotic distributions at a point where up to $(r-1)^{th}$ derivative vanishes]\label{AsymTheo2}
Suppose that $\mu$ is a convex function such that at $x_0$, $\mu^{(1)}(x_0)=\ldots =\mu^{(r-1)}(x_0)=0$ and $\mu^{(r)}(x_0)\neq 0$. Convexity of $\mu$ shows that $\mu^{(r)}(x_0) >0 $ and that $r \ge 2$ is even. We further assume that $\mu^{(r)}$ is continuous in a neighborhood around $x_0$. Then for the LSE $\hm$ it follows that 
\begin{eqnarray*}
\left(\begin{array}{l}
n^{\frac{r}{2r+1}}d_1(r,\mu)\left(\hm(x_0)-\mu(x_0)\right)\\n^{\frac{r-1}{2r+1}}d_2(r,\mu)\left(\hm^{\prime}(x_0)-\mu^{\prime}(x_0)\right)
\end{array}\right) \overset{d}{\to} \left(\begin{array}{l}
H^{\prime\prime}_{(r)}(0)\\H^{\prime\prime\prime}_{(r)}(0)
\end{array}\right)
\end{eqnarray*}
where $(H^{\prime\prime}_{(r)}(0), H^{\prime\prime\prime}_{(r)}(0))$ are the second and third derivatives at $0$ of the invelope $H_{(r)}$ of $Y_{(r)}$ (as defined in Theorem~\ref{AsymTheo1}) and 
\[d_1(r,\mu)=\left(\frac{(r+2)!}{\sigma^{2r+2}\mu^{(r)}(x_0)}\right)^{\frac{1}{2r+1}}\quad\text{and}\quad d_2(r,\mu)=\left(\frac{((r+2)!)^3}{\sigma^{2r}\left(\mu^{(r)}(x_0)\right)^3}\right)^{\frac{1}{2r+1}}.\]
\et
\begin{remark}\label{AsymTheo2remark}
 In fact Theorem~\ref{AsymTheo2} can be strengthened to show that suitably scaled version of $(\hat \mu_n,\hat \mu_n')$ (locally) converges in distribution to the stochastic process $(H^{\prime\prime}_{(r)}, H^{\prime\prime\prime}_{(r)})$ in the metric of uniform convergence on compacta, i.e.,
 \begin{eqnarray}\label{eq:AsympTheoGen}
\left(\begin{array}{l}
n^{\frac{r}{2r+1}}d_1(r,\mu)\left(\hm(x_0+tn^{\frac{-1}{2r+1}})-\mu(x_0)\right)\\n^{\frac{r-1}{2r+1}}d_2(r,\mu)\left(\hm^{\prime}(x_0+tn^{\frac{-1}{2r+1}})-\mu^{\prime}(x_0)\right)
\end{array}\right)\overset{d}{\to} \left(\begin{array}{l}
H^{\prime\prime}_{(r)}(t)\\H^{\prime\prime\prime}_{(r)}(t)
\end{array}\right)
\end{eqnarray}
where $(H^{\prime\prime}_{(r)}(t), H^{\prime\prime\prime}_{(r)}(t))$ are the second and third derivatives at $t$ of the invelope $H_{(r)}$ of $Y_{(r)}$ and $d_1(r,\mu)$ and $d_2(r,\mu)$ are defined in Theorem~\ref{AsymTheo2}.
\end{remark}
Next we study the behavior of $\hat \mu_n$ and $\hat \mu_n'$ under scenario (b), i.e., when $\mu$ is affine in an interval around $x_0$. 
 \bt\label{AsymTheo3}
Let $X(t)=W(t)$, where $W(t)$ is a standard Brownian motion on the interval $[0,1]$, and let $Y$ be the integral of $X$, satisfying $Y(0)=0$, i.e., $Y(t)=\int_0^t X(t) dt$, for $0\leq t\leq 1$. Then there exists an a.s.~uniquely defined random continuously differentiable function $H$ satisfying the following conditions:
\be
\ii  $H$ is always above the function $Y$ in the interval $[0,1]$, i.e.,
\[H(t)\geq Y(t) \quad\text{for each $t\in[0,1]$},\]
\ii  $H$ has a convex second derivative, and with probability 1, $H$ is  three times differentiable on $[0,1]$,
\ii $H(0)=Y(0)$, $H(1)=Y(1)$, $H^\prime(0)=Y^\prime(0)$, $H^\prime(1)=Y^\prime(1)$,
\ii  $H$ satisfies 
\[\int_{0}^1\left\{H(t)-Y(t)\right\}dH^{\prime\prime\prime}(t)=0.\]
\ee
\et
\begin{remark}\label{AsymTheo3remark}
One can find a detailed proof of the above theorem in~\citep[Theorem 3.5]{CW15}. The proof of the above result is essentially the same as that of Theorem~\ref{AsymTheo1} except for the fact that there are some extra boundary conditions that need to be taken care of. 
\end{remark}
\bt[Asymptotic distributions at a point where $\mu$ is affine]\label{AsymTheo4}
Suppose that $\mu$ is a convex function such that $\mu(x)=mx+c$ for some $m,c\in \RR$, in the interval $[a,b]$ around $x_0$. Then for the LSE $\hm$ over the set of all convex functions, it follows that 
\begin{eqnarray*}
\left(\begin{array}{l}
\sqrt{n}\left(\hm(x_0)-\mu(x_0)\right)\\\sqrt{n}\left(\hm^{\prime}(x_0)-\mu^{\prime}(x_0)\right)
\end{array}\right) \stackrel{d}{\to} \sigma \left(\begin{array}{l}
H^{\prime\prime}\left(\frac{x_0-a}{b-a}\right)\\H^{\prime\prime\prime}\left(\frac{x_0-a}{b-a}\right)
\end{array}\right)
\end{eqnarray*}
where $(H^{\prime\prime}(t), H^{\prime\prime\prime}(t))$ are the second and third derivatives at $t$ of the invelope $H$ of $Y$. 
\et

\begin{remark}\label{AsymTheo4remark}
Theorem~\ref{AsymTheo4} can be further strengthened to show that $(\hat{\mu}_n,\hat{\mu}^\prime_n)$ converges on any interval $[a-\delta_1,b+\delta_2]$, for $\delta_1+\delta_2 < b-a$, uniformly to the process $(H^{\prime\prime},H^{\prime\prime\prime})(\frac{\cdot - a}{b - a})$. For a proof of the Theorem \ref{AsymTheo4}, we refer to~\citep[Theorem 3.5]{CW15} where, indeed, the stronger version has been proved. Note that in~\citep{CW15} the authors consider a uniform fixed design setup whereas we use a random design setting. Thus, suitable modifications are required to apply~\citep[Theorem 3.5]{CW15} to derive our result. A key step in this regard is the uniform tightness result proved in Theorem~\ref{Ratetheo2}. 
\end{remark}
\subsection{Proof of the theorems stated in this section}

\subsubsection{Proof of the Theorem \ref{AsymTheo2}}
At first we introduce some notation. As earlier, denote the piecewise linear function through the points $(x_{i},\hm(x_i))$ by $\hm:[0,1]\to \RR$. Let us  define 
\begin{align*}
\SS_n(t)&:=\frac{1}{n}\sum_{i=1}^n Y_i\mathbf{1}_{x_i\leq t},\\
\RR_n(t)&:=\frac{1}{n}\sum_{i=1}^n \hm(x_i)\mathbf{1}_{x_i\leq t}=\int_0^t\hm(s)dF_n(s),\\
\tilde{\RR}_n(t)&:=\int_0^t\hm(s)ds,
\end{align*}
where $F_n$ is the empirical distribution function of the $x_i$'s.
Similar to that in \citep{GJW2001b}, we define the processes
\[Y_n(x) :=\int_0^x\SS_n(v)dv,\quad H_n(x) :=\int_0^x\RR_n(v)dv,\quad \tilde{H}_n(x) :=\int_0^x\tilde{\RR}_n(v)dv,\] 
as well as
\begin{align*}
Y^{loc}_n(t)& :=n^{\frac{r+2}{2r+1}}\int_{x_0}^{x_0+tn^{-\frac{1}{2r+1}}}\left\{\SS_n(v)-\SS_n(x_0)
-\vphantom{=}\mu(x_0)\int_{x_0}^v dF_n(u)\right\}dv,\\
\Hl(t)&:=n^{\frac{r+2}{2r+1}}\int_{x_0}^{x_0+tn^{-\frac{1}{2r+1}}}\left\{\Rn(v)-\Rn(x_0)-\vphantom{=}\mu(x_0)\int_{x_0}^v dF_n(u)\right\}dv+A_nt+B_n,\\
\tHl(t)&:=n^{\frac{r+2}{2r+1}}\int_{x_0}^{x_0+tn^{-\frac{1}{2r+1}}}\left\{\tRn(v)-\tRn(x_0)-\vphantom{=}\mu(x_0)\int_{x_0}^v du\right\}dv+A_nt+B_n,
\end{align*}
where 
\[A_n :=n^{\frac{r+1}{2r+1}}\left(\Rn(x_0)-\Sn(x_0)\right)\quad\text{and}\quad  B_n :=n^{\frac{r+1}{2r+1}}
\left(H_n(x_0)-Y_n(x_0)\right).\]
As in \citep{GJW2001b}, we have the following expressions:
\begin{align*}
(\tHl)^{\prime\prime}(t) &=n^{\frac{r}{2r+1}}\left(\hm(x_0+tn^{-\frac{1}{2r+1}})-\mu(x_0)\right),\\
(\tHl)^{\prime\prime\prime}(t) &=n^{\frac{r-1}{2r+1}}\left\{\hm^{\prime}(x_0+tn^{-\frac{1}{2r+1}})\right\}.
\end{align*}
From Lemma~\ref{CharLemma2}, we see that
\begin{align*}
H_n(x)\geq Y_n(x)\quad\forall \; x\in [0,1]
\end{align*}  
and equalities hold at the kinks. By virtue of Lemma~\ref{CharLemma2}, it is also easy to observe that
\[A_n=n^{\frac{r+1}{2r+1}}\left\{\Rn(x_0)-\Rn(U_n)-\left(\Sn(x_0)-\Sn(U_n)\right)\right\},\]
where $U_n:=\sup\{x\in\Omega:x\leq x_0\}$ .

 Using the results in Theorem~\ref{Ratetheo1} the tightness of the sequence $\{A_n\}$ and $\{B_n\}$ can be easily derived; the arguments will be similar to the proof given in \citep[p.~1693]{GJW2001b}.
 Now,
\begin{align*}
\Hl(t)-\Yl(t)&=n^{\frac{r+2}{2r+1}}\int_{x_0}^{x_0+tn^{-\frac{1}{2r+1}}}\left\{\Rn(u)-\Rn(x_0)-\left(\Sn(u)-\Sn(x_0)\right)\right\}du+A_nt+B_n\\
&=n^{\frac{r+2}{2r+1}}\int_{x_0}^{x_0+tn^{-\frac{1}{2r+1}}}\left\{\Rn(u)-\Sn(u)\right\}du+B_n\\
&=n^{\frac{r+2}{2r+1}}\left(H_n(x_0+n^{-\frac{1}{2r+1}})-Y_n(x_0+n^{-\frac{1}{2r+1}})\right)\geq 0.
\end{align*}
 We will state an important proposition on weak convergence of the stochastic process $\Yl$, in the metric of uniform convergence on compacta. 
\bp\label{prop:limitdist1}
We have 
\[\Yl(t) \stackrel{d}{\to} \sigma\int_0^t W(s)ds+\frac{1}{(r+2)!}\mu^{(r)}(x_0)t^{r+2}\]
in the metric of uniform convergence on compacta. 
\ep

Now~\eqref{eq:Conv1} along with the result in \eqref{eq:ImpRes1} imply that $\Hl$ and $\tHl$ are asymptotically the same. The arguments, again, are similar to those in~\citep[p.~1696]{GJW2001b}. We scale $\Hl$ properly to obtain a new process $H^{l}_n$ and use the same scaling to $\Yl$ to obtain $Y^{l}_n$. Let $k_1\Yl(k_2t)$ be the transformation that makes it converge to an integrated Brownian motion with a drift of $t^{r+2}$. Note that by the scaling property of Brownian motion we have that $\alpha^{-1/2}W(\alpha t)$ is a standard Brownian motion for all $\alpha>0$ if $W$ is one. Thus, it is easy to see that choosing 
 \[k_1=\left((r+2)!\right)^{-\frac{3}{2r+1}}\sigma^{-\frac{2r+4}{2r+1}}\left(\mu^{(r)}(x_0\right))^{\frac{3}{2r+1}}\quad\text{and}\quad k_2=\left((r+2)!\right)^{\frac{2}{2r+1}}\sigma^{\frac{2}{2r+1}}\left(\mu^{(r)}(x_0)\right)^{-\frac{2}{2r+1}}\] 
 will make $Y^{l}_n \stackrel{d}{\to} Y_{(r)}$, in the metric of uniform convergence on compacta. Further, note that
 \[(H^{l}_n)^{\prime\prime}(t)=k_1k_2^2(\tHl)^{\prime\prime}(t)=n^{\frac{r}{2r+1}}d_1(r,\mu)\left(\hm(x_0+tn^{-\frac{1}{2r+1}})-\mu(x_0)\right),\]
 and
 \[(H^{l}_n)^{\prime\prime\prime}(t)=k_1k_2^3(\tHl)^{\prime\prime\prime}(t)=n^{\frac{r-1}{2r+1}}d_2(r,\mu) \hm^{\prime}(x_0+tn^{-\frac{1}{2r+1}}).\]
The proof will now be complete if we can show that $H^{l}_n$ converges in such a way that the second and third derivatives of this invelope converges in distribution to the corresponding quantities of $H_{(r)}$ mentioned in the statement of the theorem. For proving that, we use similar arguments as in~\citep{GJW2001b}. Let us define for $c>0$ the product space $\mathcal{E}[-c,c]$ as follows:
 \[\mathcal{E}[-c,c]:=(\mathcal{C}[-c,c])^4\times (\mathcal{D}[-c,c])^2\] and endow $\mathcal{E}[-c,c]$ with the product topology induced by the uniform topology on $\mathcal{C}[-c,c]$ and the Skorohod topology on $\mathcal{D}[-c,c]$. The space $\mathcal{E}[-c,c]$ supports the vector-valued stochastic process \[ Z_n :=\left\{\left(H^{l}_n,(H^l_n)^{\prime},(H^l_n)^{\prime\prime},Y^{l}_n,(H^l_n)^{\prime\prime\prime},(Y^{l}_n)^{\prime}\right)\right\}.\]
It may be noted that the subset of $\mathcal{D}[-c,c]$ consisting of all nondecreasing functions, absolutely
bounded by $M < \infty$, is compact in the Skorohod topology. Hence, Theorem~\ref{Ratetheo1} together with the monotonicity of $(H^l_n)^{\prime\prime\prime}$ shows that the sequence $(H^l_n)^{\prime\prime\prime}$ is tight in $\mathcal{D}[-c,c]$, endowed with the Skorohod topology. Moreover, as the set of continuous functions with its values as well as its derivatives absolutely bounded by $M$ is compact in $\mathcal{C}[-c,c]$, under the uniform topology, the sequences
$H^l_n$, $(H^l_n)^{\prime}$ and $(H^l_n)^{\prime\prime}$
are also tight in $\mathcal{C}[-c,c]$. This follows from Theorem \ref{Ratetheo1}.
Since $Y^{l}_n$ and $(Y^{l}_n)^\prime$
 both converge weakly, they are also tight in $\mathcal{C}[-c,c]$ and
$\mathcal{D}[-c,c]$ respectively. This means that for each $\epsilon>0$ we can construct a compact product set in $\mathcal{E}[-c,c]$ such that the vector $Z_n$ will be contained in that set with probability at least $1-\epsilon$ for all $n$. This means that the sequence $Z_n$ is tight in $\mathcal{E}[-c,c]$.
Fix an arbitrary subsequence $\{Z_{n^{\prime}}\}$. Then we can construct a subsequence
$\{Z_{n^{\prime\prime}}\}$ such that $\{Z_{n^{\prime\prime}}\}$ converges weakly to some $Z_0$ in $\mathcal{E}[-c,c]$, for each
$c>0$. By the continuous mapping theorem, it follows that the limit $Z_0=(H_0,H^\prime_0,H^{\prime\prime}_0,Y_0,H^{\prime\prime\prime}_0)$ satisfies both
\begin{equation}\label{eq:lastAsym1}
\underset{t \in [-c,c]}{\inf}\left(H_0(t)-Y_0(t)\right)\geq 0 \quad \text{for each} \p\p c>0
\end{equation}
 and
 \begin{equation}\label{eq:lastAsym2}
\int_{[-c,c]}\{H_0(t)-Y(t)\}dH^{\prime\prime\prime}_0(t)=0
\end{equation}
a.s. Inequality \eqref{eq:lastAsym1} can, for example, be seen by using convergence of expectations of the nonpositive continuous function $\tau:\mathcal{E}[-c,c]\to \RR$ defined
by \[\tau(z_1,z_2,\ldots ,z_6)=\inf_t \left(z_1(t)-z_4(t)\right)\wedge 0.\]
Note that $\tau(Z_n)\equiv 0$ a.s. This gives $\tau(Z_0)=0$ a.s., and hence \eqref{eq:lastAsym1}. Note also
that $H^{\prime\prime}_0$ is convex and decreasing. The equality \eqref{eq:lastAsym2} follows from considering
the function
\[\tau(z_1,z_2,\ldots ,z_6)=\int_{[-c,c]}\left(z_1(t)-z_4(t)\right)dz_5(t),\]
which is continuous on the subset of $\mathcal{E}[-c,c]$ consisting of functions with $z_5$ increasing. Now, since $Z_0$ satisfies \eqref{eq:lastAsym1} for all $c>0$, and for $Y_0=Y_{(r)}$ as defined in Theorem \ref{AsymTheo1}, we see that the first condition of Theorem \ref{AsymTheo1} is satisfied by the
first and fourth components of $Z_0$. Moreover, the second condition also holds true for $Z_0$.
 Hence it follows that the limit $Z_0$ is in fact equal to $Z=(H_{(r)},H^\prime_{(r)},H^{\prime\prime}_{(r)},Y_{(r)},H^{\prime\prime\prime}_{(r)},Y^\prime_{(r)})$ involving the unique function $H_{(r)}$ described in Theorem \ref{AsymTheo1}. Since the limit for any such subsequence is the same in the uniform topology of compacta, it follows that the full sequence $\{Z_n\}$ converges weakly and has the same limit, namely $Z$. In particular $Z_d(0)\to_d Z(0)$, and this yields the result of Theorem \ref{AsymTheo2}.

\section{Behavior of the LSE at the boundary}\label{Boundary}
In the following lemma, behavior of LSE at the two boundary points, namely 0 and 1, will be studied. We show that the LSE is inconsistent at the boundary. In fact, the derivative of the LSE is unboundedness at the boundary.

As we have seen before, the LSE $\hat \mu$ is defined uniquely only at the data points $x_i$s. We use the following linear interpolation to define it at the boundary point 0:
\begin{equation}\label{eq:mu}
	\hm(0):=\hm(x_1)-\frac{\hm(x_2)-\hm(x_1)}{x_2-x_1}x_1\quad\text{and}\quad\hm^{\prime}(0):=\frac{\hm(x_2)-\hm(x_1)}{x_2-x_1}
\end{equation}
\bl\label{IncoUnbdLem1}
Suppose that $\mu$ is decreasing at $0$ and $\mu(0)>0$. Then the following hold.
\bei
\ii There exists $\epsilon_0 >0$  such that for all $\epsilon \leq \epsilon_0$,
\begin{equation}\label{eq:incounbd1}
	\liminf_{n \to \infty} \PP \left(\hm(0)>(1+\epsilon)\mu(0)\right)>0.
\end{equation}
This shows that $\hm(0)$ is inconsistent in estimating $\mu(0)$.
\ii For some $C>0$, depending only on $\epsilon_0$, for all $M>0$,
\begin{equation}\label{eq:incounbd2}
\liminf P\left(|\hm^{\prime}(0)|>M\right)>C.
\end{equation}
Thus, $\hm^{\prime}(0)$ is unbounded in probability.

Suppose now that $\mu$ is increasing at $1$ and $\mu(1)>0$. Then~\eqref{eq:incounbd1} and~\eqref{eq:incounbd2} hold which $\hm(0)$ and $\hm^{\prime}(0)$ replaced by $\hm(1)$ and $\hm^{\prime}(1)$, defined as
\[\hm(1):=\hm(x_n)+\frac{\hm(x_n)-\hm(x_{n-1})}{x_n-x_{n-1}}(1-x_n)\quad\text{and}\quad\hm^{\prime}(1):=\frac{\hm(x_n)-\hm(x_{n-1})}{x_n-x_{n-1}}.\]
\ee
\el

In case of $\mu$ being non-decreasing at $0$ or $\mu(0)\leq 0$, the same result as in Lemma \ref{IncoUnbdLem1} can also be proved by using equivariance property summarized in Lemma \ref{equivariance}.

 \section{Estimation of the point of minimum and its asymptotic distribution}\label{Minima}
In many applications it is of interest to find the point of minimum (i.e., argmin) of a regression function; see e.g.,~\cite{MR2007268} and the references therein. Intuitively, one can argue that the argmin of the estimated regression function can serve as an estimator of the true argmin. Since our estimated convex regression function is piecewise affine, it is not hard to see that the argmin for the estimated function will be one of the kink points. So, a natural estimator for the argmin of $\mu$ is 
\begin{equation}\label{eq:defmin}
  \hat{\psi}_{n,\min}= \underset{X_i}{\arg \min}\left\{\hm(X_i): 1\leq i\leq n\right\}.
\end{equation}
If argmin of $\left\{\hm(X_i): 1\leq i\leq n\right\}$ is not unique, then define $\hat{\psi}_{n,min}$ to be the minimum among all argmins.
\bt\label{InfCon} 
Let $\mu_{\min} := \arg \min_{x \in [0,1]} \mu(x)$ be the unique argmin of the true convex regression function $\mu$ and that $\mu_{\min} \in (0,1)$. Then \[\hat{\psi}_{n,\min}\overset{a.s}{\to}\mu_{\min}.\] 
\et 
\begin{remark}
The above result would also hold if $\mu_{\min} \in \{0,1\}$. A proof of this could be obtained by using an extension of~\cite[Corollary 1]{DFJ04} to the random design setup.
\end{remark}
Next we study the rate of convergence of $\hat{\psi}_{n,\min}$. It can be noted that if $\mu$ is very flat near $\mu_{\min}$ then it is difficult to estimate the exact argmin. This is because Theorem~\ref{Ratetheo1} implies that with increasing flatness of $\mu$ (i.e., increasing $r$) around $\mu_{\min}$, $\hm$ gets flatter in a wider neighborhood around $\mu_{\min}$ (as $\hat{\mu}'_n$ gets closer to $0$). Thus, it is expected that $\hat{\psi}_{n,\min}$ (which should be at one of the end points of an affine segment of $\hm$) gets further away from $\mu_{\min}$. We state a result that not only gives the rate of convergence of $\hat{\psi}_{n,\min}$ but also provides its asymptotic distribution, under mild smoothness conditions. One can note that~\cite[Theorem~3.6]{MR2509075} also gives a similar result in the case of log-concave density estimation. 
\bt\label{InfAsymp}
Let $\mu_{min} \equiv x_0 \in (0,1)$ be the unique argmin such that $\mu^{(1)}(x_0)=\ldots =\mu^{(r-1)}(x_0)=0$
 and $\mu^{(r)}(x_0)\neq 0$ where $r\ge 2$ is even. Let us assume further that $\mu^{(r)}(x_0)$ is continuous in a neighborhood around $x_0$. Then under assumption (\textbf{A1}), $\hat{\psi}_{n,\min}$ satisfies the following: 
 \begin{align}\label{eq:minrate1}
 n^{\frac{1}{2r+1}}\left(\hat{\psi}_{n,\min}-\mu_{\min}\right)\overset{d}{\rightarrow}\frac{1}{d_1(r,\mu)} \arg \min_{t\in\RR}H^{\prime\prime}_{(r)}(t),
 \end{align}
where $H''_{(r)}$ and $d_1(r,\mu)$ are defined in Theorems~\ref{AsymTheo1} and~\ref{AsymTheo2} respectively. 
\et

\appendix  
 \section{Proofs}\label{Appendix}
\subsection{Proof of Proposition \ref{Pr:Representation}}\label{pf:Prop2.1} 
We know that for a given $t\in \RR$, $(x-t)_{+}$ is a convex function. Hence it is easy to see from Lemma~\ref{CharLemma1} that 
 \begin{equation}\label{eq:char}
 \sum_{i=1}^n\left(x_i-t\right)_{+}\left(Y_i-\hm(x_i)\right)\leq 0\quad \mbox{ for all } t\in \RR.
 \end{equation}
When $t$ belongs to the set of kinks of $\hm$, $\psi(x) = \hm(x) + \delta (x - t)_+$ is a convex function, for $|\delta|$ small enough. Thus, by Lemma~\ref{CharLemma1}, we know that equality occurs in~\eqref{eq:char} when $t$ belongs to the set of kinks of $\hm$. From \citep[Lemma 2.6]{GJW2001b} we know that $\hm(x)$ can be represented as 
\begin{equation}\label{eq:OldRepr}
 \hm(x)=\hat{a}+b_0x+\sum_{j=1}^k b_j(x-x_{m_j})_{+}
\end{equation}
for some $b_j>0$ for $j=1,\ldots ,k$. Now if we plug-in the above form of $\hm$ in~\eqref{eq:char} with $t$ being a kink of $\hm$ (so that we have equality in~\eqref{eq:char}) we get the normal equations for the least squares problem in~\eqref{eq:ParaLSE}. Hence the result follows. \qed

\subsection{Proof of Lemma \ref{equivariance}} \label{pf:Lem2.3} 
It is sufficient to show that $\hat{\psi}_n $ in~\eqref{eq:equiveq} satisfies all three conditions in Lemma~\ref{CharLemma1} with $Y_i$ replaced by $Y_i+a+bx_i$. Since an affine function is always convex, condition $(i)$ follows immediately. It is also easy to see that, for any $a,b \in \R$,
   \[\sum_{i=1}^n\left(a+bx_i\right)\left(Y_i-\hm(x_i)\right)=0.\]  
This verifies condition $(ii)$ in Lemma~\ref{CharLemma1}.  Condition $(iii)$ in Lemma~\ref{CharLemma1} can also be verified similarly.\qed


 \subsection{Proof of Lemma \ref{1stSteplemma}}

 We know that for all $x\in \RR$ and $\delta>0$, $\psi(t)=\delta(t-x)_{+}$ is a convex function. As the sum of two convex functions is also convex, $\hm(t)+\delta(t-x)_{+}$ will also be convex for all
$\delta \ge 0$. If $\hm$ has a change in slope at $x$, for some sufficiently small $\delta>0$, $\hm(t)-\delta(t-x)
 _{+}$ will also be a convex function. Now if we plug $\psi(t)=\hm(t)+ (t-x)_{+}$ for all $x\in[0,1]$ and 
 $\psi(t)=\hm(t)-\delta(t-x)_{+}$, for $x$ being a point of slope change of $\hm$, into the third condition of 
Lemma~\ref{CharLemma1}, we obtain the desired result. \qed

 \subsection{Proof of Lemma \ref{2ndSteplemma}}
 
 Let us fix any $x$ from $T$. Along with $x$, let us choose its left nearest sample point and right
 nearest sample point and call them $a_x$ and $b_x$, respectively. From~\eqref{eq:Theo1} we know that $G(a_x)\geq 0$ and $G(b_x)\geq 0$. Now, as $G(x) = 0$, we have
 \[G(b_x)-G(x)=(b_x-x)\sum_{x_i\leq x}\left(Y_i-\hat{\mu}_n(x_i)\right)\geq 0\] 
 and \[G(a_x)-G(x)=(a_x-x)\sum_{x_i\leq a_x}\left(Y_i-\hat{\mu}_n(x_i)\right)\geq 0.\]
Thus, $\sum_{x_i\leq x} (Y_i-\hat{\mu}_n(x_i)) \ge 0$ and $\sum_{x_i\leq x} (Y_i-\hat{\mu}_n(x_i)) - (Y_m-\hat{\mu}_n(x)) \le 0$, where $m$ is the index of $x$, i.e, $x_m=x$. These two facts imply that  
\[|\sum_{x_i\leq x}\left(Y_i-\hat{\mu}_n(x_i)\right)|\leq \left|Y_m-\hat{\mu}_n(x)\right|.\]
 Now if we can bound $\underset{x\in T}{\sup}\left|Y_m-\hat{\mu}_n(x)\right|$,
 we are done. Applying triangle inequality, we have
\begin{equation}\label{eq:SplitIn2}
	\underset{x_i\in T}{\sup}\left|Y_i-\hat{\mu}_n(x_i)\right|\leq \underset{x_i\in [a+\epsilon,b-\epsilon]}{\sup}\left|\epsilon_i\right|+\underset{x\in [a-   \epsilon,b+\epsilon]}{\sup}\left|\hat{\mu}_n(x)-\mu(x)\right|.
\end{equation}
The first term in the right hand side of~\eqref{eq:SplitIn2} is $O_p(\log n)$ thanks to assumption (\textbf{A1}). As $\hm$ uniformly converges to $\mu$ a.s.~in any compact interval completely contained in the interior of the support of $X$, the second term in the right hand side of~\eqref{eq:SplitIn2} is $o_p(1)$. Hence the result follows. \qed

\subsection{Proof of Lemma \ref{3rdSteplemma}}

$(i)$ Choose $\delta>0$ such that $\hat{\mu}_n(x)+\delta f_{u,v}(x)$ is a convex function. Note that this is possible as $u$ and $v$ are kinks. Now if we use $\hat{\mu}_n(x)+ \delta f_{u,v}(x)$ as the convex function $\psi$ in the characterization Lemma~\ref{CharLemma1}$(iii)$ we obtain the desired result. \newline

\noindent $(ii)$  Observe that 
\begin{align*}
Z(u,v)-\frac{1}{n}\sum_{u\leq x_i\leq v}f_{u,v}(x_i)Y_i &=\frac{1}{n}\sum_{x_i < u \textrm{ or }  x_i> v}\left(Y_i-\hm(x_i)\right)-\frac{1}{n}\sum_{u \le x_i \le v }f_{u,v}(x_i)\hm(x_i).
\end{align*} 
Noticing that $\sum_{x_i> v}\left(Y_i-\hm(x_i)\right) = - \sum_{x_i \le v}\left(Y_i-\hm(x_i)\right)$, and using~\eqref{eq:Theo2} we can see that the supremum of the first term (on the right hand side of the above display) over the set $T$ is
 $O_p(n^{-1}\log n)$. As $\hm(x)$ will be affine in the interval $[u,v]$, for $u$ and $v$ being consecutive kinks, we can further expand the second term as 
 \[\frac{1}{n}\sum_{u \le x_i \leq v }f_{u,v}(x_i)\hm(x_i)=\frac{1}{n}\sum_{u \leq x_i \leq v }f_{u,v}(x_i)\left\{\hm(u)+\hm^{\prime}(u+)(x_i-u)\right\}.\]

Next, we will show that 
\begin{equation*}
\sqrt{\frac{n}{V_n-U_n}} \max\left\{ \Big| \frac{1}{n}\sum_{i:U_n\leq x_i\leq V_n}f_{U_n,V_n}(x_i)(x_i-U_n)\Big|, \Big| \frac{1}{n}\sum_{i:U_n\leq x_i\leq V_n}f_{U_n,V_n}(x_i)\Big| \right\}=O_p(1),
\end{equation*} 
which will complete the proof, as $\hm(U_n)$ and $\hm^{\prime}(U_n+)$ are $O_p(1)$. Fix $K_1, K_2 >0$. Choose $\delta_n >0$ such that $K_1c_n < \delta_n < K_2c_n$. For any $\delta >0$, let us define the following two function classes:
 \[\mathfrak{F}^{(1)}_{\delta}=\left\{\frac{f_{u,v}(x)\mathbf{1}_{[u,v]}(x)}{\sqrt{v-u}} : x_0-\delta\leq u\leq
  x_0-K_1c_n \leq x_0+K_1c_n\leq v\leq x_0+\delta\right\}\]and\[ \mathfrak{F}^{(2)}_{\delta}=\left\{
  \frac{f_{u,v}(x) \mathbf{1}_{[u,v]}(x)(x-u)}{\sqrt{v-u}} : x_0-\delta\leq u\leq x_0-K_1c_n  \leq x_0+K_1
  c_n \leq v\leq x_0+\delta\right\}.\]
We can take 
 \[F^{(1)}_{\delta}(x) :=\frac{2\mathbf{1}_{[x_0-\delta,x_0+\delta]}(x)}{\sqrt{2K_1c_n}}\quad\text{and}\quad F^{(2)}
 _{\delta}(x):=\frac{2\mathbf{1}_{[x_0-\delta,x_0+\delta]}(x)\delta}{\sqrt{2K_1c_n}}\]
as the envelope functions for the classes $\mathfrak{F}^{(1)}_{\delta}$ and $\mathfrak{F}^{(2)}_{\delta}$, respectively. Using Theorem 2.14.1 of \citep{VAN&WELL2000}, we get  
\begin{eqnarray*}
\mathbb{E}\left[\left(\underset{g\in \mathfrak{F}^{(1)}_{\delta_n}}{\sup}\left|(\mathbb{P}_n-P)g\right| \right)^2\right] & \leq & \frac{K}{n} \mathbb{E}\left[F^{(1)}_{\delta}(X)\right]^2=O(n^{-1}), \;\; \mbox{and}  \\ 
\mathbb{E}\left[\left(\underset{g\in \mathfrak{F}^{(2)}_{\delta_n}}{\sup}\left|(\mathbb{P}_n-P)g\right|\right)^2\right]
& \leq & \frac{K}{n} \mathbb{E}\left[F^{(2)}_{\delta}(X)\right]^2 = O(n^{-1}c_n^{2}),
\end{eqnarray*}
where $K$ is some constant.

From~\eqref{eq:V_n-U_n}, for any given $\epsilon>0$,  there exist $K_1, K_2>0$ such that the event $A_n := \{K_1c_n\leq V_n-U_n\leq K_2c_n\}$ has  probability at least $1-\epsilon$ for all $n$. So
\begin{eqnarray*}
&& \mathbb{P} \left(\sqrt{\frac{n}{V_n-U_n}}\Big| \frac{1}{n}\sum_{i:U_n\leq x_i\leq V_n}f_{U_n,V_n}(x_i)(x_i-U_n)\Big|>M\right)\\
 &\leq & \mathbb{P}\left(\sqrt{\frac{n}{V_n-U_n}}\Big|\frac{1}{n}\sum_{i:U_n\leq x_i\leq V_n}f_{U_n,V_n}(x_i)(x_i-U_n)\Big|>M, A_n \right) + \PP \left(A_n^c\right) \\
&\leq & \mathbb{P}\Big (\sqrt{n}\underset{g\in \mathfrak{F}^{(2)}_{\delta_n}}{\sup}\left|(\mathbb{P}_n-P)g\right|>M \Big)+\epsilon\\
  &\leq & \frac{n}{M^2} \mathbb{E}\Big[\Big(\underset{g\in \mathfrak{F}^{(2)}_{\delta_n}}{\sup}\left|(\mathbb{P}_n-P)g\right|\Big)^2\Big]+\epsilon \leq \frac{O(c_n^{2})}{M^2}+\epsilon. 
 \end{eqnarray*}
 Similarly we can show that, using the function class $\mathfrak{F}^{(1)}_{\delta_n}$,
 \begin{align*}
 \mathbb{P}\left(\sqrt{\frac{n}{V_n-U_n}}\left| \frac{1}{n}\sum_{i:U_n\leq x_i\leq V_n}f_{U_n,V_n}(x_i)\right|>M\right)
 \leq \frac{C}{M^2}+\epsilon 
 \end{align*}
 for some constant $C>0$. As $\{c_n\}^{\infty}_{n=1}$ is bounded we can make both the tail probabilities as small as possible by choosing $M$ sufficiently large. 
\qed

\subsection{Proof of Proposition \ref{Prop:1st}}
Let us denote by $I := \{(u,v) \in [0,1]^2: x_0 - \delta_0 <u<x_0<v < x_0 + \delta_0\}$. Define the random variable $M_n$ to be the smallest value for which~\eqref{eq:Pn-P} holds for all $(u,v) \in I$. Define $g_{u,v}(x) := f_{u,v}(x) \mathbf{1}_{[u,v]}(x) (x-u)^r$, for $x \in (0,1)$. 
Then $$\PP(M_n>m) = \PP\Big(\mbox{there exists } (u,v) \in I: \left|\left(\mathbb{P}_n-P\right)g_{u,v}(X)\right|>\epsilon(v-u)^{r+1}+n^{-(r+1)}m^2\Big)$$ can be bounded from above by {\small
\begin{eqnarray*}
&& \sum_{j\in \NN:jn^{-1}<2 \delta_0} \PP\left(\exists \; (u,v) \in I, \frac{j-1}{n}\leq v-u\leq \frac{j}{n}:\left|\left(\mathbb{P}_n-P\right)g_{u,v}(X)\right|>\epsilon(v-u)^{r+1}+n^{-(r+1)}m^2\right)\\
&\leq & \sum_{j\in \NN:jn^{-1}<2 \delta_0}\PP\left(\exists \; (u,v) \in I, \frac{j-1}{n}\leq v-u\leq \frac{j}{n}: n^{r+1}\left|\left(\mathbb{P}_n-P\right)g_{u,v}(X)\right|>\epsilon(j-1)^{r+1}+m^2\right)\\
&\leq & c^{\prime} \sum_{j\in \NN:jn^{-1}<2 \delta_0} n^{2r+2}\dfrac{\EE \Big[\Big(\underset{g\in \mathfrak{F}_{jn^{-1}}}{\sup}\left|(\mathbb{P}_n-P)g\right|\Big)^2\Big]}{\{\epsilon(j-1)^{r+1}+m^2\}^2}\\
&\leq &\sum_{j\in \NN:jn^{-1}<2 \delta_0} c^{\prime\prime}\dfrac{j^{2r+1}}{\{\epsilon(j-1)^{r+1}+m^2\}^2}
\end{eqnarray*}}
where the last inequality follows from~\eqref{eq:Impone}. Therefore, we can ensure that the sum is arbitrarily small, by choosing $m$ large enough. \qed

\subsection{Proof of Lemma \ref{Lem:1stResult1}}

Fix $\delta_0>0$. Let us define $I_\delta:=\{(u,v)\in\RR^2:x_0-\delta\le u\le x_0-{\delta_0}\le x_0+\delta_0\le v\le x_0+\delta\}$. Consider the following class of functions:
\[\mathcal{F}_\delta(x):=\left\{\frac{f_{u,v}(x)\mathbf{1}_{[u,v]}(x)\varepsilon}{\sqrt{v-u}}:(u,v)\in I_\delta\right\}.\]
Note that $F_\delta(x) :=\mathbf{1}_{[u,v]}(x)|\varepsilon|/\sqrt{2\delta_0}$ can be taken as an envelop function for the above class of functions. It can be observed that $\EE[F^2_\delta(X)]\lesssim\delta\sigma^2/\delta_0$. Recall that for sufficiently large $n$, there exists $0<K_1<K_2$ such that $K_1c_n<V_n-U_n<K_2c_n$ with probability at least $1-\epsilon$ for some bounded sequence of positive real numbers $\{c_n\}^{\infty}_{n=1}$. In particular if we set $\delta_0=K_1c_n$, then $\EE[F^2_{K_2c_n}(X)]\lesssim K_2\sigma^2/K_1$. Using Theorem 2.14.1 of \citep{VAN&WELL2000}, we have
\begin{eqnarray*}
\PP\left(\sqrt{\frac{n}{V_n-U_n}}\left| Z_2(U_n,V_n)\right|>M\right)&\leq & \PP\left(\sup_{g\in \mathcal{F}_{K_2c_n}}\sqrt{n}\left|(\PP_n-P)g\right|>M\right)+\epsilon\\
&\le & \frac{n\EE\Big[\Big(\sup_{g\in \mathcal{F}_{K_2c_n}}\left|(\PP_n-P)g\right|\Big)^2\Big]}{M^2}+\epsilon \\
&\lesssim & \frac{\EE[F^2_{K_2c_n}(X)]}{M^2}+\epsilon \lesssim \frac{K_2\sigma^2}{K_1M^2}+\epsilon.
\end{eqnarray*}
Thus, the above tail probability can be made arbitrarily small by choosing $M$
sufficiently large. Hence,
\begin{equation}\label{eq:Z_2rate}
Z_2(U_n,V_n)=O_p\left(\sqrt{\frac{V_n-U_n}{n}}\right).
\end{equation}
  Lemma~\ref{3rdSteplemma} shows that $Z(U_n,V_n)\leq 0$. Let us consider the expansion of $Z(U_n,V_n)$ as shown in \eqref{eq:expanRate}. From \eqref{eq:Z_1rate}, it can be seen that the leading term in the expansion of $Z_1(U_n,V_n)$ is nonnegative and 
   is  $O_p((V_n-U_n)^{r+1})$, whereas $Z_2(U_n,V_n)=O_p(n^{-1/2}\sqrt{V_n-U_n})$. This enforces 
  \[\left(V_n-U_n\right)^{r+1}\lesssim \sqrt{\frac{V_n-U_n}{n}}\]
   which yields  $V_n-U_n=O_p(n^{-1/(2r+1)})$. \qed

 \subsection{Proof of Lemma \ref{Lem:1stResult2}}
 
 The following expansion can be obtained from the definitions of $G(\mathcal{U}_n)$ and $G(\mathcal{V}_n)$
 \begin{align}\label{eq:expan2}
 0=G(\mathcal{V}_n)-G(\mathcal{U}_n)&=\left(\mathcal{V}_n-\mathcal{U}_n\right)\sum_{x_i\leq \mathcal{U}_n}\left(Y_i-\hat{\mu}_n(x_i)\right)+\sum_{\mathcal{U}_n<x_i\leq \mathcal{V}_n}\left(\mathcal{V}_n-x_i\right)\varepsilon_i\nonumber\\
 &\phantom{=}+\sum_{\mathcal{U}_n<x_i\leq \mathcal{V}_n}\left(\mathcal{V}_n-x_i\right)\left(\mu\left(x_i\right)-\hat{\mu}_n(x_i)\right).
 \end{align}
 It follows from the given conditions and Lemma~\ref{2ndSteplemma} that the first term in \eqref{eq:expan2} is  $O_p(n^{-1/(2r+1)}\log n)$ . Using similar  technique used in determining the rate of convergence of $Z_2(U_n, V_n)$ in Lemma~\ref{Lem:1stResult1}, one can see
 \begin{equation}\label{eq:2ndtermorderIN1stResult2}
 \sum_{\mathcal{U}_n<x_i\leq \mathcal{V}_n}\left(\mathcal{V}_n-x_i\right)\varepsilon_i=O_p\left(n^{1/2}\left(\mathcal{V}_n-\mathcal{U}_n\right)^{3/2}\right).
 \end{equation}
 
  We have to prove that given any $\epsilon>0$, there exists $c_\epsilon>0$ such that 
 $\inf_{\mathcal{U}_n\leq x\leq \mathcal{V}_n}|\hat{\mu}_n(x)-\mu(x)|<c_\epsilon n^{-\frac{r}{2r+1}}$
  with probability greater than $1-\epsilon$. Let us suppose that this is not the case.  So there exists $c_n\uparrow \infty$ such that $\inf_{\mathcal{U}_n\leq x\leq \mathcal{V}_n}\left|\hat{\mu}_n(x)-\mu(x)\right|>c_n n^{-r/(2r+1)}$ holds with probability greater than $\epsilon$. Continuity of $\hm(x)-\mu(x)$ implies that $\hm(x)-\mu(x)$ are of same sign for all $x$ in the interval $[\mathcal{U}_n,\mathcal{V}_n]$ whenever $\inf_{\mathcal{U}_n\leq x\leq \mathcal{V}_n}|\hat{\mu}_n(x)-\mu(x)|>0$. So according to our assumption 
 \begin{equation}\label{eq:2ndterm1orderIN1stResult2}\left|\sum_{\mathcal{U}_n<x_i\leq \mathcal{V}_n}\left(\mathcal{V}_n-x_i\right)\left(\mu\left(x_i\right)-\hat{\mu}_n(x_i)\right)\right|\gtrsim c_n n^{1-\frac{2}{2r+1}-\frac{r}{2r+1}} =c_n n^{\frac{1}{2}-\frac{3}{2(2r+1)}}
 \end{equation} holds also with a probability greater than at least $\epsilon/2$ for all sufficiently large $n$.
It follows from Lemma~\ref{Lem:1stResult1} and \eqref{eq:2ndtermorderIN1stResult2} that the second term in right hand side of \eqref{eq:expan2} is $O_p\left(n^{\frac{1}{2}
 -\frac{3}{2(2r+1)}}\right)$. That contradicts the equality in \eqref{eq:expan2}.

 Hence \[\underset{\mathcal{U}_n\leq x\leq \mathcal{V}_n}{\inf}\left|\hat{\mu}_n(x)-\mu(x)\right|=O_p\left(n^{-\frac{r}{2r+1}}\right).\] \qed


\subsection{Proof of Lemma \ref{difflemma}}
Let $\hm(x)$ satisfies
\[\hm(x)=p+qx\]
for all $x\in [u,v]$ for some $p,q \in \RR$. It is easy to see that $\hat{a}_{ls}+\hat{b}_{ls}x=\bar{Y}+\hat{b}_{ls}(x-\bar{x})$ where $\bar{Y}$ denotes mean of the observation in the interval $[u,v]$. So we can write down the following
\[\left|\hat{a}_{ls}+\hat{b}_{ls}x-\hm(x)\right|\leq \left|x-\bar{x}\right|\left|\hat{b}_{ls}-q\right|+\left|\bar{Y}-p-q\bar{x}\right|.\] 
Now it follows from the definition of $\t$ in \ref{Shortdef} that 
\[\left|\bar{Y}-p-q\bar{x}\right|=\left|\frac{\t}{k_2-k_1+1}\right|.\]
On the other hand it can be also observed that
\[\tt-\bar{x}\t=\sum_{i: u\leq x_i\leq v}\left(x_i-\bar{x}\right)\left(Y_i-qx_i\right)=\left[\sum_{i: u\leq x_i\leq v}(x_i-\bar{x})^2\right]\left(\hat{b}_{ls}-q\right).\]
Now $|x-\bar{x}|\leq |v-u|$ for all $x\in [u,v]$. Hence the result follows.

\subsection{Proof of Lemma \ref{Hingelemma}}
To get an upper bound on $\sup_{x\in [u,v]}\left|\hat{a}_{ls}+\hat{b}_{ls}x-\hm(x)\right|$ one can note from Lemma~\ref{difflemma} that it suffices to prove some upper bounds on $\t$, $\tt$ and lower bounds on $\sum_{i: u\leq x_i\leq v}(x_i-\bar{x})^2$ and $k_2-k_1+1$. Upper bounds on $\t,\tt$ can be essentially traced back to~\eqref{eq:imp8} and \eqref{eq:imp9}, respectively. Let us now recall that in any compact set $\mathfrak{C}$ completely contained in $(0,1)$ the following holds:
\begin{equation}\label{eq:linear5theq}
\sup_{x_i \in \mathfrak{C}}|Y_i-\hm(x_i)|=O_p\left(\log n\right).
\end{equation}
This is a consequence of the sub-gaussian assumption on the error distribution and the uniform convergence of $\hm$ to $\mu$ on $\mathfrak{C}$. Consequently, one can see from \eqref{eq:linear5theq} that the right side of \eqref{eq:imp8} and \eqref{eq:imp9} are $O_p(\log n)$. Also, $1/(k_2-k_1+1)$ and $1/(\sum_{i: u\leq x_i\leq v}(x_i-\bar{x})^2)$ are $O_p(1/n)$, thanks to the assumption $1/(v-u)=O_p(1)$. Combining all those results, we have 
\begin{equation}\label{eq:linear6theq(rep)}\sup_{x\in [u,v]}\left|\hat{a}_{ls}+\hat{b}_{ls}x-\hm(x)\right|=O_p\left(\frac{\log n}{n}\right).
\end{equation}

\subsection{Proof of Proposition~\ref{prop:limitdist1}}
To prove the result, we will break $\Yl(t)$ into three parts, similar to what was done in \citep[Theorem 6.2]{GJW2001a}, and show the convergence of each part separately. Observe that,
\begin{eqnarray*}
\Yl(t) &=&n^{\frac{r+2}{2r+1}}\int_{x_0}^{x_0+tn^{-\frac{1}{2r+1}}}\left\{\Sn(v)-\Sn(x_0)-\int_{x_0}^v \mu(x_0) dF_n(u)\right\}dv\\
&=&n^{\frac{r+2}{2r+1}}\int_{x_0}^{x_0+tn^{-\frac{1}{2r+1}}}\left\{\Sn(v)-\Sn(x_0)-\left(R_0(v)-R_0(x_0)\right)\right\}dv\\&\phantom{=}&+ \; n^{\frac{r+2}{2r+1}}\int_{x_0}^{x_0+tn^{-\frac{1}{2r+1}}}\left\{R_0(v)-R_0(x_0)-\int_{x_0}^v\mu(x_0)dF_n(u)\right\}dv\\
&=&n^{\frac{r+2}{2r+1}}\int_{x_0}^{x_0+tn^{-\frac{1}{2r+1}}}\left\{\frac{1}{n}\sum_{i:x_0<x_i\leq v}\epsilon_i\right\}dv\\
\end{eqnarray*}
\begin{eqnarray*}
\phantom{\Yl(t)}&\phantom{=}&+ \; n^{\frac{r+2}{2r+1}}\int_{x_0}^{x_0+tn^{-\frac{1}{2r+1}}}\left\{\frac{1}{n}\sum_{i:x_0<x_i\leq v}\mu(x_i)-\int_{x_0}^v\mu(u)du\right\}dv\\
&\phantom{=}&- \; n^{\frac{r+2}{2r+1}}\int_{x_0}^{x_0+tn^{-\frac{1}{2r+1}}}\left\{\mu(x_0)\int_{x_0}^v d(F_n(u)-u)\right\}dv \\
&\phantom{=}&+ \; n^{\frac{r+2}{2r+1}}\int_{x_0}^{x_0+tn^{-\frac{1}{2r+1}}}\left\{R_0(v)-R_0(x_0)-\mu(x_0)\int_{x_0}^v du\right\}dv\\
&=&{\mathbf{I}}_n(t)+ {\mathbf{II}}_n(t) +{\mathbf{III}}_n(t)
\end{eqnarray*}
where 
\begin{align}
\mathbf{I}_n(t) &:= n^{\frac{r+2}{2r+1}}\int_{x_0}^{x_0+tn^{-\frac{1}{2r+1}}}\left\{\frac{1}{n}\sum_{i:x_0<x_i\leq v}\epsilon_i\right\}dv\label{eq:Expr1},\\
\mathbf{II}_n(t) &:= n^{\frac{r+2}{2r+1}}\int_{x_0}^{x_0+tn^{-\frac{1}{2r+1}}}\left\{\frac{1}{n}\sum_{i:x_0<x_i\leq v}\mu(x_i)-\int_{x_0}^v\mu(u)du\right\}dv\nonumber\\&\phantom{=} -n^{\frac{r+2}{2r+1}}\int_{x_0}^{x_0+tn^{-\frac{1}{2r+1}}}\left\{\mu(x_0)\int_{x_0}^vd(F_n(u)-u)\right\}dv \label{eq:Expr2},\\
\mathbf{III}_n(t) &:= n^{\frac{r+2}{2r+1}}\int_{x_0}^{x_0+tn^{-\frac{1}{2r+1}}}\left\{R_0(v)-R_0(x_0)-\mu(x_0)\int_{x_0}^v du\right\}dv.\label{eq:Expr3}
\end{align}
Let us start with $\mathbf{I}_n(t)$:
\begin{align*}
\mathbf{I}_n(t)&= n^{\frac{r+2}{2r+1}}\int_{x_0}^{x_0+tn^{-\frac{1}{2r+1}}}\left\{\frac{1}{n}\sum_{i:x_0<x_i\leq v}\epsilon_i\right\}dv = n^{-\frac{r-1}{2r+1}}\sum_{i:x_0<x_i\leq x_0+tn^{-\frac{1}{2r+1}}}\epsilon_i(x_0+tn^{-\frac{1}{2r+1}}-x_i).
\end{align*}
Let us define $\mathbf{X}=(X_1,\ldots , X_n)$. Note that $\mathbb{E}(\mathbf{I}_n(t)|\mathbf{X})=0$. Let us apply the transformation $t_i=n^{\frac{1}{2r+1}}(x_i-x_0)$ and define $G_n(\cdot)$ as 
\begin{equation}\label{eq:ImpRes1}
G_n(x) :=n^{\frac{1}{2r+1}}\left(F_n(x_0+xn^{-\frac{1}{2r+1}})-x_0\right).
\end{equation}
As $\{\mathbf{1}(\cdot\leq x):x\in \RR\}$ is a Glivenko-Cantelli class of function, $G_n(x)$ converges uniformly to $x$ a.s., for all $|x|\leq c$ for every $c>0$. Hence \[\mbox{Var}\left(\mathbf{I}_n(t)\vert \mathbf{X}\right)=\sigma^2{n^{-\frac{2r}{2r+1}}}\sum_{i:0<t_i\leq t}(t-t_i)^2=\sigma^2\int_0^t(t-x)^2dG_n(x)\stackrel{a.s.}{\rightarrow} \frac{\sigma^2}{3} t^3.\]
Thus,
\[\mbox{Var}\left(\mathbf{I}_n(t)\right)=\mathbb{E}\left[\mbox{Var}\left(\mathbf{I}_n(t)\vert {\mathbf{X}}
\right)\right]\to \frac{\sigma^2}{3} t^3\]
which is the variance of $\sigma^2\int_0^t W(s)ds$. Using Theorem 2.11.1 of \citep{VAN&WELL2000}, it can be noted that $\mathbf{I}_n(t)$ converges weakly to $\sigma^2\int_0^tW(s)ds$, uniformly on compacta.

 Now we consider $\mathbf{II}_n(\cdot)$. Notice that 
\begin{eqnarray}
\mathbf{II}_n(t) &=& n^{\frac{r+2}{2r+1}}\int_{x_0}^{x_0+tn^{-1/(2r+1)}}\left(\int_{(x_0,v]}(\mu(u)-\mu(x_0))d(F_n(u)-u)\right)dv\nonumber\\
&=& n^{\frac{r+2}{2r+1}}\int_{x_0}^{x_0+tn^{-1/(2r+1)}}\left(\int_{(x_0,v]}\frac{\mu^{(r)}(w_u)}{r!}(u-x_0)^r d(F_n(u)-u)\right)dv\nonumber\\
&=& n^{\frac{r+1}{2r+1}}\int_{0}^{t}\left(\int_{x_0}^{x_0 + v'n^{-1/(2r+1)}}\frac{\mu^{(r)}(w_u)}{r!}(u-x_0)^r d(F_n(u)-u)\right)dv'\nonumber\\
&=& \int_0^t\left(\int_{(0,v^\prime]}\frac{\mu^{(r)}(w_{u'})}{r!}(u')^r (dG_n(u')-du^\prime)\right)dv^\prime,
\end{eqnarray} 
 where $v \mapsto v' := (v - x_0) n^{-1/(2r+1)}$, $u \mapsto u' := (u-x_0) n^{-1/(2r+1)}$, $w_u$ is an intermediate point between $u$ and $x_0$. Due to a.s.~uniform convergence of $G_n(x)$ to $x$ on $[-c,c]$ for some $c>0$, it can be shown that $\mathbf{II}_n(\cdot)$ will converge to $0$ uniformly (a.s.) on $[-c,c]$. Now let us consider $\mathbf{III}_n(\cdot)$. Observe that the deterministic term $\mathbf{III}_n$ can be simplified as:
  \begin{align*}
 \mathbf{III}_n(t)&=n^{\frac{r+2}{2r+1}}\int_{x_0}^{x_0+tn^{-\frac{1}{2r+1}}}\left\{R_0(v)-R_0(x_0)-\mu(x_0)\int_{x_0}^v du\right\}dv\\
 &=n^{\frac{r+2}{2r+1}}\int_{x_0}^{x_0+tn^{-\frac{1}{2r+1}}}\frac{1}{(r+1)!}\mu^{(r)}(x_0)(v-x_0)^{r+1}dv+o(1) \to \frac{1}{(r+2)!}\mu^{(r)}(x_0)t^{r+2}, 
\end{align*}  
uniformly on compacta. This completes the proof. \qed

\subsection{Proof of Lemma \ref{IncoUnbdLem1}}
At first we will show that $(i)$ implies $(ii)$. Then proof of $(i)$ will be given later.
 
Fix $M>0$ and $\epsilon <\epsilon_0$. From the convexity of $\hm$ it follows that for any $t>0$, $\hm(0)\leq \hm(t)-\hm^{\prime}(0)t$. Since $\mu$ is decreasing near 
$0$, hence $\hm^{\prime}(0)<0$ a.s. So we have $\hm(0)\leq \hm(t)+\left|\hm^{\prime}(0)\right|t$. Note that,
\begin{align}\label{eq:incoUnbd3}
   \PP\left(\hm(0)>(1+\epsilon)\mu(0)\right)\leq \PP\left(\hm(t)>\left(1+\frac{\epsilon}{2}\right)\mu(0)\right)+ \PP\left(|\hm^{\prime}(0)|>\frac{\epsilon}{2t}\mu(0)\right).
\end{align}
Taking liminf on both side, we have
\begin{align}\label{eq:incoUnbd4}
  \liminf_{n \to \infty} \PP\left(\hm(0)>(1+\epsilon)\mu(0)\right)&\leq \liminf_{n \to \infty} \PP\left(\hm(t)>\left(1+\frac{\epsilon}{2}\right)\mu(0)\right)\\&\phantom{=} \;\;\;+\liminf_{n \to \infty} \PP\left(|\hm^{\prime}(0)|>\frac{\epsilon}{2t}\mu(0)\right).\nonumber
   \end{align} 
Let us choose $t$ such that $\mu(0)\epsilon \ge 2Mt$ and $\mu(t) \le (1+\epsilon/2)\mu(0)$, e.g., $t$ can be taken as
   \[t:= \frac{\mu(0)\epsilon}{2M}\wedge \left(1+\frac{\epsilon}{2}\right)\mu(0).\]
For such choice of $t$, we get \[\liminf_{n \to \infty} \PP\left(\hm(t)>\left(1+\frac{\epsilon}{2}\right)\mu(0)\right)=0.\]
Using $(i)$, we can say  
   \[\liminf_{n \to \infty} \PP\left(|\hm^{\prime}(0)|>M\right) \ge \liminf \PP\left(\hm(0)\left(1+\epsilon\right)\mu(0)\right)>0.\]
  
 Next we show that $(i)$ holds. Let us recall the first inequality in Lemma \ref{CharLemma2}:
\begin{align}
   \hm(x_1)&\geq Y_1.\label{eq:incoUnbd5}
\end{align}
   Since $\mu$ is decreasing near zero, hence $\hm^{\prime}(0)<0$ a.s. Hence from~\eqref{eq:incoUnbd5}, we get that 
\begin{eqnarray*}   
\hm(0)-\mu(0) & > & Y_1 - \mu(x_1) + (\mu(x_1) - \mu(0)) \\ 
& \ge & \varepsilon_1+o_p(1).
\end{eqnarray*}
As $\varepsilon_1$ is a mean zero non-degenerate random variable, for some $\epsilon_0 >0$, $\PP(\varepsilon_1>\epsilon_0\mu(0)) > 0$. This completes the proof of $(i)$.
 
To prove the analogous results for $\hm(1)$ observe that $\hm(1)\geq Y_n$. This follows from Lemma~\ref{CharLemma2} by subtracting the equality in~\eqref{eq:charlemma2} for $j=n$ from  the inequality for $j=n-1$. The rest of the proof is similar. \qed

\subsection{Proof of Theorem \ref{InfCon}}

Given $\epsilon >0$, we will show that $\limsup_{n\to\infty}|\hat{\psi}_{n,\min} - \mu_{\min}|<\epsilon$ a.s. That will imply our result because we can vary $\epsilon$ over the set of positive rationals. So for proving the claim, let us fix an $\epsilon>0$ such that $[\mu_{\min}-\epsilon/2,\mu_{\min}+\epsilon/2] \subset (0,1)$. As $\mu_{\min}$ is assumed to be the unique argmin, $$\delta :=\min \left\{-\lim_{x \downarrow \mu_{\min}-\epsilon/2} \mu^{\prime}(x),\lim_{x \uparrow \mu_{\min}+\epsilon/2} \mu^{\prime}(x) \right\}>0.$$ From the uniform convergence of $\hat \mu'_n$ to $\mu'$ on compacts within $(0,1)$ (see e.g.,~\citep[Lemma 5]{MAMMEN1991}, \citep[Theorem 3.2]{SS11}) it follows that for sufficiently large $n$, $\lim_{x \downarrow \mu_{\min}-\epsilon/2} \hat{\mu}_n^{\prime}(x) <-\delta/2$ and $\lim_{x \downarrow \mu_{\min}+\epsilon/2} \hat{\mu}_n^{\prime}(x) >\delta/2$ would occur a.s. That shows the a.s.~existence of the argmin of $\hat{\mu}_n$ around the $\epsilon$-neighborhood of $\mu_{\min}$, for sufficiently large $n$. Hence the claim follows. \qed  
 
 \subsection{Proof of Theorem \ref{InfAsymp}}
For notational convenience we write $\mu_{\min} \equiv x_0$.
Note that the result in~\eqref{eq:AsympTheoGen} in Remark~\ref{AsymTheo2remark} holds for  $\mu_{\min} \equiv x_0$.


We can show that Lemma 7 of \citep{GJW2001b} can be generalized in a way that for a given $\epsilon>0$ there exists $M=M(\epsilon)>0$, independent of $t$, such that 
\begin{eqnarray}\label{eq:ModTight}
P\left(\left|H^{\prime\prime\prime}_{(r)}(t)-\frac{(r+2)!}{r!}t^{r-1}\right|>M\right)<\frac{\epsilon}{2}.
\end{eqnarray} 
 We can choose a sufficiently large $T>0$ so that $M<\frac{(r+2)!}{r!}T^{r-1}$. Hence, by virtue of the fact that $r$ is even
 \begin{equation}\label{er:ModTightCor}
 P\left(H^{\prime\prime\prime}_{(r)}(T)>0, H^{\prime\prime\prime}_{(r)}(-T)<0\right)>1-\epsilon.
 \end{equation}  
Now~\eqref{eq:AsympTheoGen} helps us conclude that for sufficiently large $n$ the following holds:
 \[P\left(\hm^{\prime}(x_0+Tn^{\frac{-1}{2r+1}})>0, \hm^{\prime}(x_0-Tn^{\frac{-1}{2r+1}})<0\right)>1-2\epsilon.\]  
Hence, as $\hat{\psi}_{n,\min}$ will be trapped inside $[x_0-Tn^{\frac{-1}{2r+1}}, x_0+Tn^{\frac{-1}{2r+1}}]$ with probability greater than $ 1 - 2 \epsilon$, we have
 \begin{equation}\label{eq:Tight}
 n^{\frac{1}{2r+1}}\left(\hat{\psi}_{n,\min}-x_0\right)=O_p(1).
 \end{equation}
Further, note that $n^{1/(2r+1)}(\hat{\psi}_{n,\min}-x_0)$ is the argmin of the stochastic process 
 \[n^{\frac{r}{2r+1}}\left(\hm(x_0+tn^{\frac{-1}{2r+1}})-\mu(x_0)\right)\] over $t\in \RR$. So by the argmax continuous mapping theorem (see e.g.,~\citep[Theorem 3.2.2]{VAN&WELL2000}) we have 
 \[n^{\frac{1}{2r+1}}(\hat{\psi}_{n,\min}-x_0)\overset{d}{\to}\frac{1}{d_1(r,\mu)}\arg \min_{t\in\RR}H^{\prime\prime}_{(r)}(t).\]\qed
\bibliographystyle{alpha}  
\bibliography{Reference}

\end{document}